\documentclass[floatfix,%
reprint,
superscriptaddress,
 amsmath,amssymb,
 aps,
prl
]{revtex4-2}

\usepackage{url}
\usepackage{graphicx}
\usepackage{dcolumn}
\usepackage{bm}
\usepackage[utf8]{inputenc}
\usepackage{textcomp}
\usepackage{braket}
\usepackage{svg}
\usepackage{ulem}
\usepackage{comment}
\usepackage{tikz}
\usepackage{appendix}
\begin{document}

\preprint{APS/123-QED}

\title{Experimental observation of strong field stabilization}  

\author{Anna R.\ Dardia}
\affiliation{Department of Physics, University of California, Santa Barbara, California, 93106, USA}
\author{Spencer Walker}
\thanks{Equal contribution.}
\affiliation{Department of Physics, The Ohio State University, Columbus, OH, 43210, USA}
\author{Yifei Bai}
\thanks{Equal contribution.}
\affiliation{Department of Physics, University of California, Santa Barbara, California, 93106, USA}
\author{Petros Kousis}
\affiliation{Department of Physics, University of California, Santa Barbara, California, 93106, USA}
\author{Alexandra S.\ Landsman}
\affiliation{Department of Physics, The Ohio State University, Columbus, OH, 43210, USA}
\author{David M.\ Weld}
\affiliation{Department of Physics, University of California, Santa Barbara, California, 93106, USA}
\email{weld@ucsb.edu}

\begin{abstract}
Bound quantum states such as atoms can be torn apart by strong oscillating fields. A natural expectation is that stronger fields lead to more certain destruction. In contradiction to this intuition, some theories predict a striking reversal: that as the field intensity is raised above some threshold, bound state wavefunctions can spatially bifurcate and become increasingly stable with increasing field intensity. This ``strong field stabilization'' was predicted decades ago in the context of atoms in pulsed laser fields, but has resisted experimental observation due to extreme laser intensity requirements and theoretical controversy. We report the experimental observation of strong-field stabilization of a ground state, using trapped neutral atoms to emulate the dynamics of atomic electrons in an extremely strong laser field.  We directly image the predicted wavepacket bifurcation, measure an ionization rate non-monotonic in field amplitude, and map out the regime of stabilization as a function of laser pulse parameters. We observe that stabilization persists down to surprisingly low drive frequencies, near and below the scale of the lowest-energy excitations of the bound state. These results confirm and extend a long-standing prediction of extreme quantum dynamics, and showcase a complementary tool for probing strong-field phenomena near and beyond the frontier of current laser technology.
 
\end{abstract}

\maketitle
\section{Introduction}
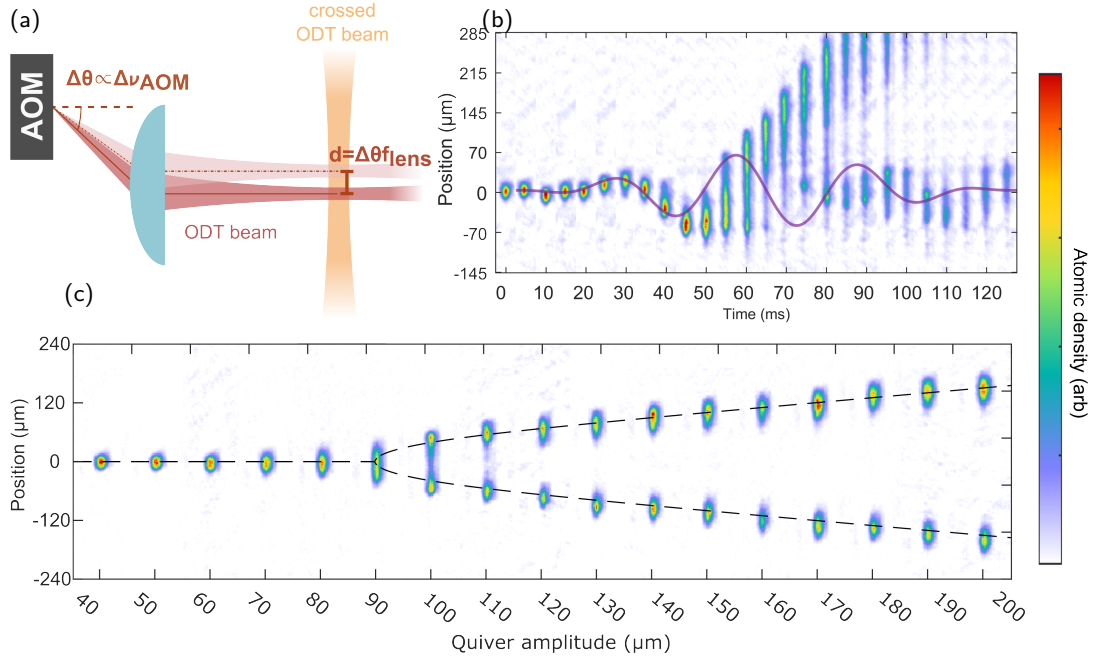
\begin{figure*}[ht!]
    \centering
    \begin{tikzpicture}
    \node[anchor=south west, inner sep=0] (img) at (0,0)
            {\includegraphics[
                width=0.8\linewidth,
            ]{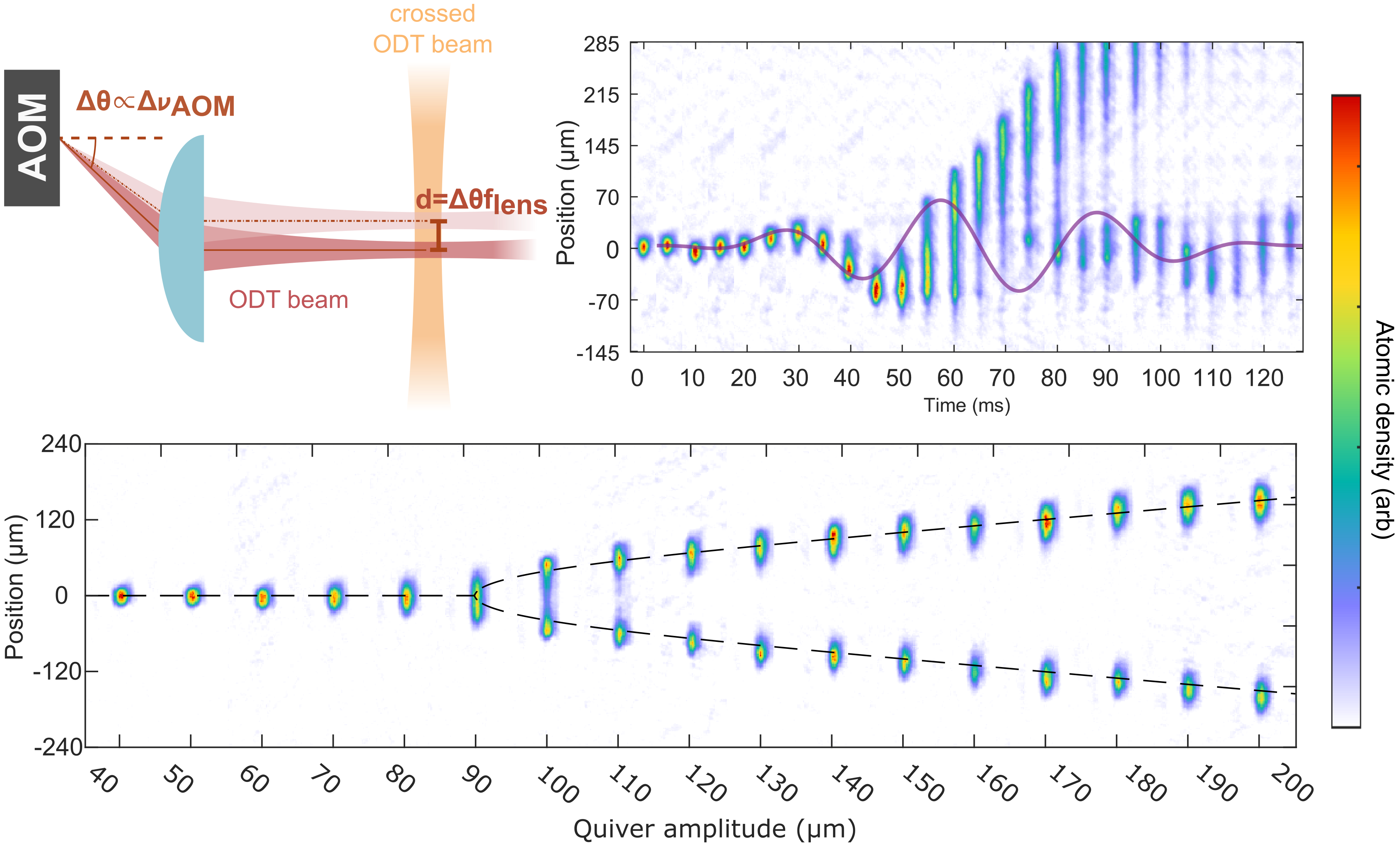}};

        \begin{scope}[x={(img.south east)}, y={(img.north west)}]
            \node[anchor=north west, xshift=-0.02\linewidth, yshift=-0.0\linewidth]
                at (0.02,1) { \textsf{(a)}};

            \node[anchor=north east, xshift=-0.02\linewidth, yshift=-0.0\linewidth]
                at (0.5,1.0) {\textsf{(b)}};

            \node[anchor=south west, xshift=0.02\linewidth, yshift=0.02\linewidth]
                at (0.02,0.48) {\textsf{(c)}};
        \end{scope}
        \end{tikzpicture}
        \caption{
        \textbf{Probing strong-field dynamics with trapped atoms}. \textbf{a)} Diagram of the experimental setup. The trap position is varied by an acousto-optic modulator to induce inertial forces which correspond to the electric fields of a laser pulse. \textbf{b)}~Absorption images of strontium atoms subjected to a 4-cycle  pulse with amplitude $A=38.3\,\mu\text{m}$ and frequency $\omega_\text{drive}=1.14\,\omega_0$ ($2\pi\times 32\,\text{Hz}$), taken at various times during the pulse to reveal sub-cycle dynamics. Solid line indicates the time-dependent position of the trap center. The observed unbinding from the trap partway through the pulse corresponds to strong-field photoionization. \textbf{c)}~Absorption images of atoms after a 250 ms ramp-up to a range of final pulse amplitudes, for a drive frequency $\omega_\text{drive}=180\,\omega_0$ ($2\pi\times5\,\text{kHz}$). A clear bifurcation or ``dichotomy'' of the wavepacket is visible above a critical amplitude. Dashed line depicts the predicted functional form of the bifurcation~\cite{SI}. Panels (b) and (c) share a common colorbar indicating atomic density in arbitrary units.}  
    \label{fig:setup}
\end{figure*}

A longstanding counter-intuitive prediction in strong-field physics proposes that sufficiently strong laser fields can stabilize atoms against ionization. 
This phenomenon, called strong-field stabilization, has been the subject of nearly 40 years of study and debate~\cite{pont_dichotomy_1988,su1990dynamics,fedorov_stabilization_1999, delone_atomic_1995, gavrila_atomic_2002, faria_stabilization_1999}.
\textcolor{black}{The origin of stabilization is the profound structural changes that atoms undergo in intense oscillating fields strong enough to rival or dominate the nuclear Coulomb field. Such fields cause quiver motion of the bound electron as though it were nearly free. In the rest frame of the electron | the Kramers-Henneberger frame | these field-induced oscillations are transferred to the nucleus; cycle-averaging in this frame results in an effective Kramers-Henneberger atomic potential~\cite{kramers1956collected, henneberger1968perturbation} which acquires a double-well structure above a critical field strength~\cite{pont_dichotomy_1988}. The bound states of this transformed energy landscape stabilize the atom against ionization, and generate the characteristic bifurcation of the electronic wavefunction.} 
While this effective potential picture appears to rely on a high-frequency approximation appropriate for UV or x-ray lasers, this description is physically appropriate whenever the time-dependent harmonics can be treated as perturbations on the time-averaged potential~\cite{popov_applicability_1999,volkova_emergence_1997,Morales_Richter_Patchkovskii_Smirnova_2011}.  The consequent possibility of stabilization at lower (e.g. optical) frequencies has been much debated in the theoretical literature~\cite{Fring1997IonizationPT, reiss_frequency_1992,popov_applicability_1999, Volkova_Popov_Tikhonova_2001, stroe_low-frequency_2008, volkova_ionization_2011, Morales_Richter_Patchkovskii_Smirnova_2011, floriani2024scars}. 
Stabilization arising from effective potentials formed by rapidly oscillating fields is reminiscent of, but distinct from, the well-known classical phenomena of the Kapitza pendulum and the Paul trap, and indeed the fundamental mechanism has general applicability: strong field stabilization has been predicted in negative ions \cite{volkova_stabilization_1994} and in otherwise unstable systems including multiply charged anions \cite{Wei_Kais_Moiseyev_2006}, positronium \cite{wei_positronium_2013}, and even molecules of bare nuclei \cite{smirnova2003molecule}.  

Direct experimental study of strong-field stabilization has been hindered by difficulty in reaching the required extreme laser field amplitudes, and also by the problem of the ``death valley"-- the region of moderate amplitudes and high ionization probability that a continuously varying laser pulse must traverse to reach the stabilization regime \cite{fedorov_stabilization_1999, gavrila_atomic_2002}. Apart from analogous phenomena studied in waveguides \cite{Longhi_Marangoni_Janner_Ramponi_Laporta_Cianci_Foglietti_2005}, experiments probing stabilization have been limited to nearly-unbound excited-state atoms driven at high frequencies \cite{de_boer_indications_1993, de_boer_adiabatic_1994, van_druten_adiabatic_1997}. The low-frequency regime remains entirely unexplored experimentally, and notwithstanding indirect evidence~\cite{eichmann_acceleration_2009}, clear observation of the \textcolor{black}{structural changes associated with stabilization} has proved elusive even in the high-frequency regime~\cite{eichmann_acceleration_2009, Morales_Richter_Patchkovskii_Smirnova_2011, wei_pursuit_2017, zimmermann_limit_2018, zhang_symmetry_2020}. 
\textcolor{black}{Classical numerics have not unambiguously settled the existence of strong field stabilization. The physical validity of the Kramers-Henneberger potential and the interpretation of the ionization rate in numerics have both been the subject of debate \cite{Geltman_1995, Geltman_Fotino_2002}; furthermore,  different numerical approaches make different sets of approximations, leading to differing predictions about the regime in which stabilization occurs if it is predicted to occur at all \cite{Geltman_1995, Geltman_Fotino_2002, geltman_short-pulse_1994, geltman_excitation_1994, geltman_volkov_1992, chen_numerical_1993, kostrykin_ionization_1997}. Only experimental results can provide the insight to resolve such debates, complementing and benchmarking differing theoretical and numerical approaches.} 

Motivated by these challenges, recent work suggests a new approach to experimental exploration of extreme dynamical phenomena like strong-field stabilization: realization of equivalent time-dependent Hamiltonians in the highly controlled context of quantum gases. Trapped ultracold atoms subjected to time-dependent forces exhibit dynamics equivalent to those of bound electrons in strong laser fields \cite{dum_wave_1998, sala_ultracold-atom_2017, Argüello-Luengo_Rivera-Dean_Stammer_Maxwell_Weld_Ciappina_Lewenstein_2024,ma2026emulationdynamicsboundelectron}, with an optical trap playing the role of the binding potential and a time-dependent force playing the role of the laser field. 
In this ultralow-energy context, attosecond-equivalent dynamics proceed on  millisecond timescales~\cite{senaratne_quantum_2018}, enabling sub-cycle time-resolved measurements of wavepacket dynamics.
In this work we probe strong-field stabilization using this approach, on an experimental platform comprising Bose-condensed $^{84}\text{Sr}$ in an optical trap subjected to inertial forces.  We observe strong-field stabilization of a ground state, demonstrate the persistence of stabilization in the low-frequency regime, directly image the predicted dichotomy of the wavepacket, quantitatively characterize the effects on stabilization of frequency and pulse shape, and explore the collective dynamics of bound and unbound wavepackets during the pulse.

\section{Experimental approach}
The experiments we report begin by Bose condensing ${N_0}\approx 250,000$ $^{84}\text{Sr}$ atoms in a crossed optical dipole trap which can be moved along one axis with an acousto-optic modulator.  Key experimental elements are depicted in Fig~\ref{fig:setup}(a) and described in more detail in the SI~\cite{SI}. The trapping potential along the direction of motion is $V_{\text{trap}}=-V_0\text{exp}(-2(x-a(t))^2/w^2)$, where $V_0$ is the trap depth, $w$ is the beam waist, and $a(t)=A(t)\text{sin}(\omega_\text{drive}t+\phi)$ describes the time-dependent displacement of the trap center, with $A$ the amplitude of motion, $\omega_\text{drive}$ the angular frequency, and $\phi$ the phase. 
In the lab frame, the shaken trap generates an inertial force equivalent to the electric-field force on a bound electron in the Kramers-Henneberger frame, with $A$ mapping to quiver amplitude \textcolor{black}{and $V_0$ and $N_0$ determining the condensate chemical potential $\lvert\mu\rvert$, which maps to the binding energy}.    In atomic units, $A$ depends on the strong laser field $E_0$ as $A=E_0/\omega_{\mathrm{drive}}^2$. 
The trap frequency $\omega_0=2\pi\times(28.0 \pm 0.3)~\text{Hz}$ characterizes the level spacing of the lowest energy states. 
Without interactions, the atomic system maps directly to a bound electron in the Kramers-Henneberger frame under the single-active electron approximation
\cite{kramers1956collected,henneberger1968perturbation,fedorov1997atomic}. An interacting cold atom system typically operates in the Thomas--Fermi regime 
\cite{pitaevskii2016bose}, similar to models used to describe static atomic properties \cite{feynman1949equations,lundqvist2013theory} and the collective electronic response of heavy atoms exposed to X-ray fields~\cite{ball1973photoabsorption}.
The role of interactions is addressed in detail in the Supplementary Information \cite{SI}(Sec. 7 and Fig. S11), which compares experimental data for different atom numbers to single-atom and mean-field numerics to demonstrate that stabilization does not depend on the presence or absence of interactions but represents a generic phenomenon of strongly driven bound systems.
As shown in Fig.~1(b), at sufficiently large quiver amplitudes atoms can undergo transitions to unbound states and be ejected from the trap, in analogy to ionization. \textcolor{black}{Strong-field stabilization would represent the suppression of this phenomenon at very strong drive amplitudes.}

This experimental context permits the exploration of extremely large drive amplitudes, at or beyond the edge of current feasibility in a pulsed laser experiment, and comparable to those at which the Kramers-Henneberger picture predicts wavefunction bifurcation. As the amplitude increases up to and beyond the regime where strong ionization is first observed, the cycle-averaged potential is predicted to acquire a double-well form above a critical amplitude $A_{\mathrm{crit}}\sim (8/9)w$~\cite{SI}.
\textcolor{black}{In hydrogen, this dichotomy was predicted to start at a field-induced quiver motion amplitude of around 20 Bohr radii~\cite{pont_dichotomy_1988}. For the VUV photons which satisfy the assumptions \textcolor{black}{of the high-frequency limit $\hbar\omega_{drive}\geq|\mu|$ in}~\cite{pont_dichotomy_1988}, this corresponds to \textcolor{black}{intensities} on the order of \textcolor{black}{$10^{19}-10^{21}~\mathrm{W/cm^2}$}, a major challenge for current laser technology in that wavelength range.}
In our experiment, for a drive slightly below the high-frequency limit with $\hbar\omega_{drive}=0.59|\mu|$, when the drive amplitude surpasses this critical value we indeed observe a clear bifurcation of the atomic density into two symmetric lobes, with a separation increasing with drive strength in quantitative agreement with theoretical predictions (Fig.~\ref{fig:setup}(c)). 
In this regime, we have verified that the bifurcation preserves quantum coherence by snapping off all traps, allowing the two bifurcated components to expand and overlap, and observing high-contrast matter-wave interference fringes between them \footnote{While more a check on the coherent nature of the bifurcation than a main subject of this work, such interferometric probes of dynamical evolution, which have also been discussed in the context of photoelectron spectra~\cite{Morales_Richter_Patchkovskii_Smirnova_2011},
represent an intriguing possible direction for further study of strong-field quantum dynamics.}.

\section{Mapping to the strong-field parameter space}

Such experiments can be characterized by a few dimensionless ratios depending only on binding energy and drive parameters~\cite{Argüello-Luengo_Rivera-Dean_Stammer_Maxwell_Weld_Ciappina_Lewenstein_2024}, allowing for direct comparison between experiments at the very different energy scales of pulsed lasers and quantum gases. Like Keldysh’s successful theory of strong-field ionization \cite{keldysh1965ionization}, such a mapping abstracts away the difference between short-range and Coulombic potentials to provide a unified picture of the dynamical regime. \textcolor{black}{The binding energy in our experiments is the field-free chemical potential of the condensate, $|\mu|\approx303~\hbar\omega_0$ for $N_0\approx250,000$ atoms}. The multiquantum parameter $K_0\equiv\lvert\mu\rvert/\hbar\omega_{\mathrm{drive}}$~\cite{v2014keldysh} expresses the binding energy in units of drive quanta, \textcolor{black}{or roughly the number of drive photons that must be absorbed to unbind the atom}. In our experiments this runs from $K_0\approx340$ \textcolor{black}{at $\omega_\mathrm{drive}=25$~Hz} to $K_0\approx1.7$ \textcolor{black}{at $\omega_\mathrm{drive}=5$~kHz} for the dichotomy results in Fig.~\ref{fig:setup}(c). The drive strength enters through the ponderomotive energy
$U_\text{p}=\tfrac14 mA^2\omega_\text{drive}^2$\textcolor{black}{, the cycle-averaged kinetic energy of the quiver motion,} and the Keldysh parameter $\gamma_\text{K}\equiv\sqrt{|\mu|/2U_\text{p}}$~\cite{keldysh1965ionization}\textcolor{black}{, which compares the quiver and binding energies}. Expressed as the laser intensity and photon energy that would reproduce the same values in a hydrogen atom (taken
as a convenient reference \textcolor{black}{with comparable binding energy to neutral atoms generally}) the effective photon energy of the trap drive $13.6\,\text{eV}/K_0$ rises from the infrared at large $K_0$ to the extreme ultraviolet as $K_0\to1$, with effective intensity $I_\text{eff}=I_\text{au}/[4(K_0\gamma_\text{K})^2]$,
where $I_\text{au}=3.51\times10^{16}\,\text{W/cm}^2$ is one atomic unit of intensity. The effective photon energies in this work span $0.04$--$8\,\text{eV}$, reaching \textcolor{black}{synthetic intensities of} $\sim10^{21}\,\text{W/cm}^2$ at the strongest drive. 

Ordinarily, $\gamma_{\mathrm{K}}\gg1$ \textcolor{black}{characterizes a perturbative field in which the atom must absorb many quanta to ionize} and $\gamma_{\mathrm{K}}\lesssim1$ characterizes \textcolor{black}{a field which bends the binding potential into a barrier through which a bound particle tunnels}. In this work the drives are much stronger. \textcolor{black}{Every drive we describe pushes past the over-the-barrier intensity $I_{\mathrm{BSI}}(\mu)$ at which the barrier is suppressed to $\mu$, the energy up to which the bound states are filled, so that the population escapes over the barrier into the continuum rather than tunneling through. Most drives are stronger still, with the inertial force overcoming the trap's maximum restoring force and removing the barrier altogether \cite{SI}, rendering the multiphoton-tunneling distinction drawn by $\gamma_\text{K}$ inapplicable in this extremely strong-field regime. } 

\begin{figure}[ht!]
    \centering
    \includegraphics[width=0.8\columnwidth]{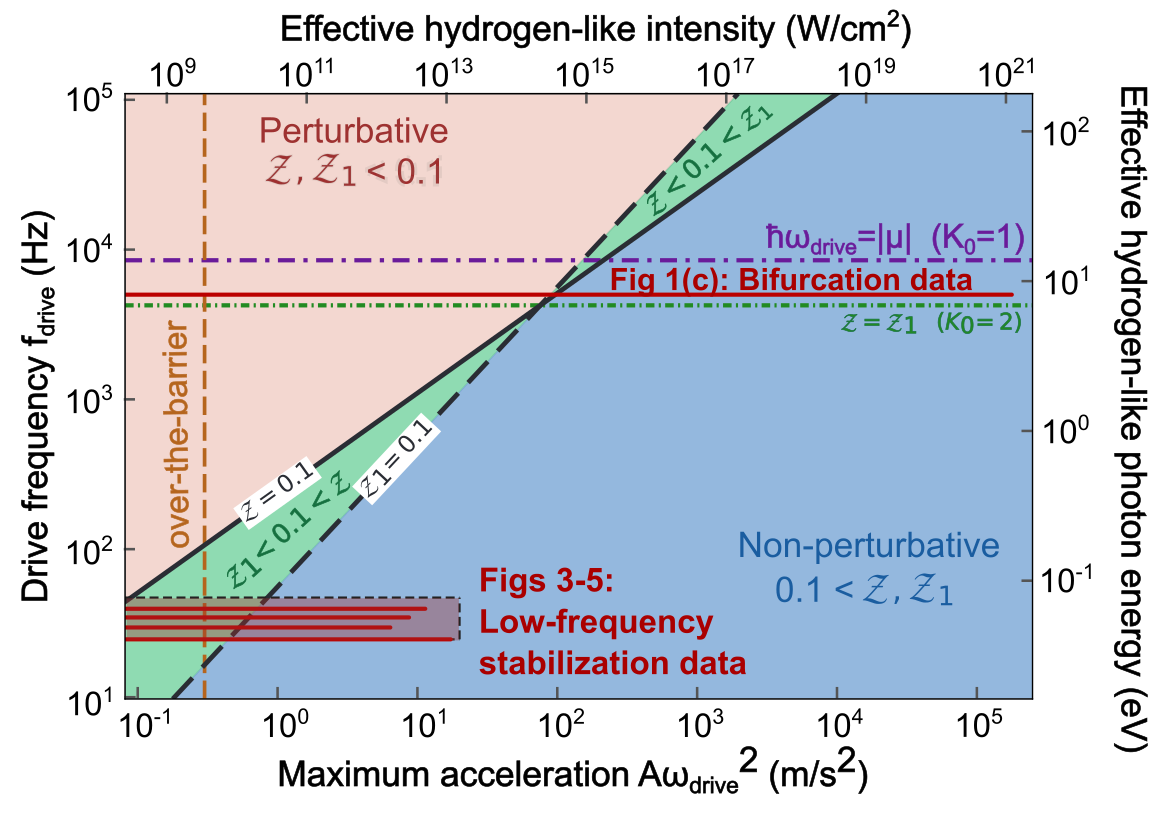}
    \caption{\textbf{Mapping the strong-field parameter space.} Drive parameters explored in these experiments. Top and right axes give effective hydrogen-referenced intensity and photon
    energy reproducing $K_0$ and $\gamma_\text{K}$
    ($|\mu|\!\leftrightarrow\!13.6$~eV). The solid $\mathcal{Z}=0.1$
    and dashed $\mathcal{Z}_1=0.1$ lines divide the plane into a perturbative
    region ($\mathcal{Z},\mathcal{Z}_1<0.1$), two intermediate bands
    ($\mathcal{Z}_1<0.1<\mathcal{Z}$ and $\mathcal{Z}<0.1<\mathcal{Z}_1$), and a
    non-perturbative region ($\mathcal{Z},\mathcal{Z}_1>0.1$). Violet and green
    dash-dotted lines mark $K_0=1$ and $K_0=2$ (where $\mathcal{Z}=\mathcal{Z}_1$);
    the orange dashed line marks the over-the-barrier intensity. Red lines denote experimental drives applied for the data shown in Figs.\ 1, 3, 4a, and 5; the shaded black box outlines the amplitude--frequency
    window over which stabilization is mapped in Fig.\ \ref{fig:PD}(b).} 
    \label{fig:params}
\end{figure}

A more natural description is in terms of Reiss's
parameters~\cite{reiss1980effect,krainov1997radiative},
$\mathcal{Z}\equiv U_\text{p}/\hbar\omega_\text{drive}$ and
$\mathcal{Z}_1\equiv2U_\text{p}/|\mu|=\gamma_\text{K}^{-2}$, which  compare the quiver energy to a single drive quantum and to the binding energy, measuring how far the drive has pushed the system out of the perturbative regime. The drive is perturbative only when both are small, $\mathcal{Z},\mathcal{Z}_1\lesssim0.1$. \textcolor{black}{Because $\mathcal{Z}/\mathcal{Z}_1=K_0/2$, }the drive frequency determines when $\mathcal{Z}$ and $\mathcal{Z}_1$ reach the non-perturbative threshold: $\mathcal{Z}$  when $K_0>2$, defining the low-frequency
strong-field limit (the focus of this work), and $\mathcal{Z}_1$ when $K_0<2$, defining the high-frequency strong-field limit, coinciding at
$K_0=2$~\cite{walker2024above}. \textcolor{black}{The dimensionless parameters are
shape-independent, but the over-the-barrier intensity is
not: }the broad, tail-less trap reaches
$I_\text{BSI}(\mu)$ at a much lower effective intensity ($\sim3\times10^9\,\text{W/cm}^2$)
than the $I_\text{BSI}(\text{H})\approx1.4\times10^{14}\,\text{W/cm}^2$ of a hydrogen
atom of equal binding energy~\cite{krainov1995energy}; had we calibrated
$I_\text{eff}(\mu)$ to \textcolor{black}{that field} rather than $\gamma_\text{K}$, every
drive would map to intensities $\sim4\times10^4$ higher, the strongest near
$\sim5\times10^{25}\,\text{W/cm}^2$. 

Figure~\ref{fig:params}, following similar diagrams in Refs.~\cite{krainov1997radiative, walker2024above}, lays out the strong-field parameter space this work explores and serves as a roadmap for the sections that follow. The drive used in Fig.~1(c), at $K_0=1.7$, is in the high-frequency strong-field limit while still somewhat below the single-photon ionization threshold. This drive is at the low-photon-energy edge of free-electron laser operation \cite{Rolles31122023} and at intensities at or beyond what those facilities can attain. The fast-ignition pulses of inertial-confinement fusion --- among the most intense fields regularly made in the laboratory at $\sim 10^{19}-10^{20}\mathrm{W/cm}^2$ \cite{betti2016inertial} --- fall in the interior of the parameter space shown in the figure. The experiments discussed in this work reach $\mathcal{Z}$ and $\mathcal{Z}_1$ $\sim10^{5}$, far into the nonperturbative regime. 
The drives of Figs.~3-5 exist in the low-frequency strong-field limit, where the drive frequency is on order $\omega_0$ and the strong-field picture is least tested. 

\begin{figure*}[ht!]
    \centering
    \begin{tikzpicture}
        \node[anchor=south west, inner sep=0] (img) at (0,0)
            {\includegraphics[
                width=0.9\linewidth,
                trim={0 230 0 230}
            ]{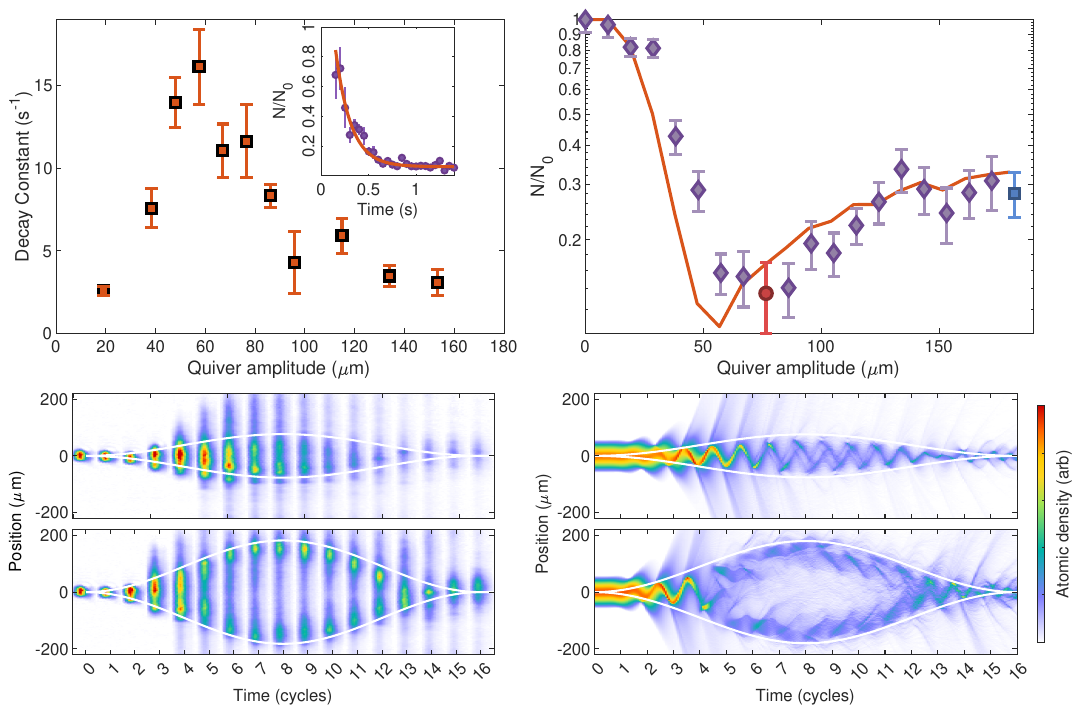}};

        \begin{scope}[x={(img.south east)}, y={(img.north west)}]

            \node[anchor=north west, xshift=0.02\linewidth, yshift=-0.02\linewidth]
                at (0.11,1) { \textsf{(a)}};

            \node[anchor=north east, xshift=-0.02\linewidth, yshift=-0.02\linewidth]
                at (0.895,1.0) {\textsf{(b)}};

            \node[anchor=south west, xshift=0.02\linewidth, yshift=0.02\linewidth]
                at (0.11,0.36) {\textsf{(c)}};

            \node[anchor=south east, xshift=-0.02\linewidth, yshift=0.02\linewidth]
                at (0.60,0.36) {\textsf{(d)}};
            \node[anchor=south west, xshift=0.02\linewidth, yshift=0.02\linewidth]
                at (0.78,0.36) {\textsf{$75.8 \mu\text{m}$}};
            \node[anchor=south west, xshift=0.02\linewidth, yshift=0.02\linewidth]
                at (0.785,0.16) {\textsf{$180\mu\text{m}$}};


            \node[anchor=south west, xshift=0.02\linewidth, yshift=0.02\linewidth]
                at (0.37,0.36) {\textsf{$76.6 \mu\text{m}$}};
            \node[anchor=south west, xshift=0.02\linewidth, yshift=0.02\linewidth]
                at (0.36,0.16) {\textsf{$181.9\mu\text{m}$}};


        \end{scope}
    \end{tikzpicture}

    \caption{\textbf{Observation of strong-field stabilization.} \textbf{a)} Measured unbinding rate in a constant amplitude field of frequency $1.43\,\omega_{\mathrm{0}}$ (40~Hz), plotted versus quiver amplitude. After a 6-cycle ramp into a flat-top pulse, the surviving population is measured via absorption imaging after varying durations, and the decay constant is determined by an exponential fit. Note the sharp decrease in loss rate above a critical quiver amplitude. Error bars denote $95\%$ confidence intervals; inset depicts one such fit. \textbf{b)} Surviving population fraction following a 16-cycle $1.43\,\omega_{\mathrm{0}}$ (40~Hz) $\text{sin}^2$-envelope pulse as a function of quiver amplitude. Symbols denote experimental data; solid red curve shows the results of numerics described in the text. Note the growth in surviving population fraction above a critical quiver amplitude. Error bars represent standard error. \textbf{c)} Experimental absorption images of time-resolved dynamics at the pulse amplitudes indicated by the red and blue markers in panel b, below and above the bifurcation threshold respectively. 
\textbf{d)} Numerical simulation of density evolution for pulse amplitudes closely matched to those in panel c.
White curves in both panels trace the pulse envelope, highlighting densities confined within the region bounded by $x=\pm A(t)$. Both panels share a common colorbar indicating atomic density.} 
    \label{fig:stabilization}
\end{figure*}

\begin{figure*}[ht]
    \centering 
        \begin{tikzpicture}
        \node[anchor=south west, inner sep=0] (img) at (0,0)
            {\includegraphics[width=0.8\linewidth]{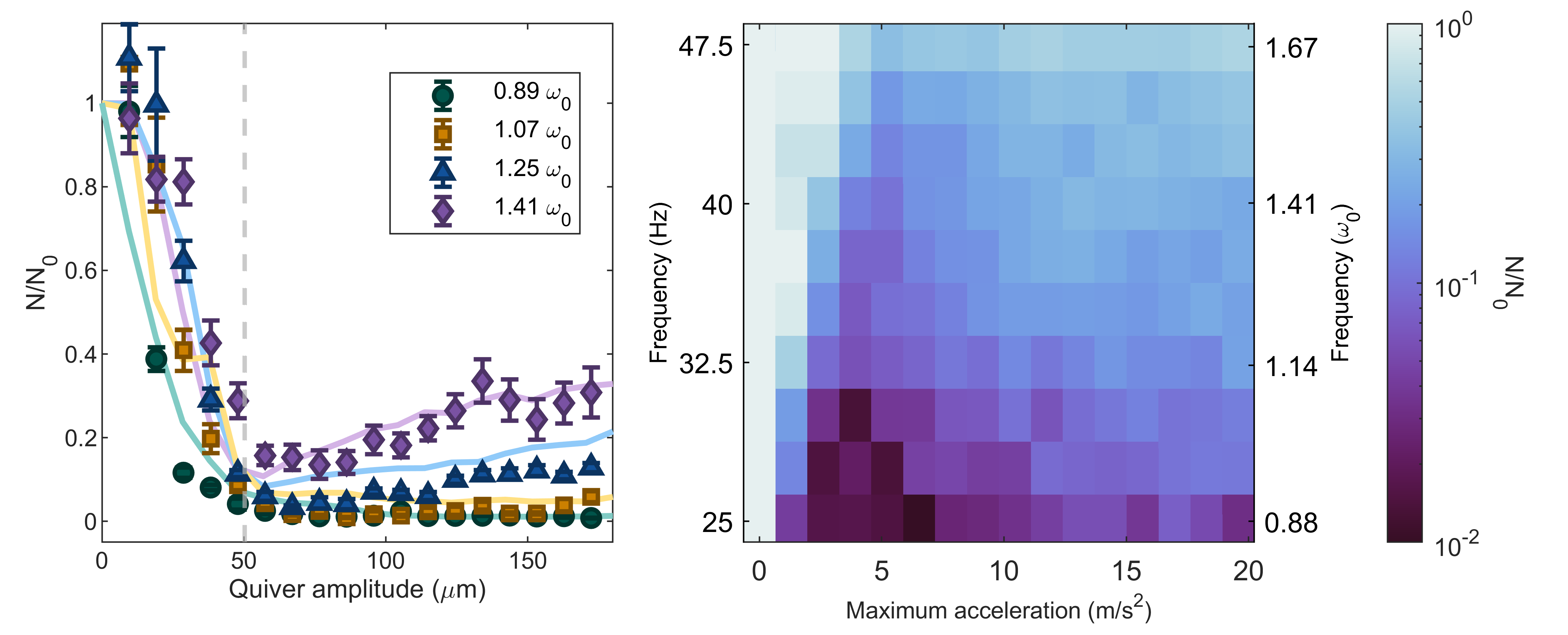}};

        \begin{scope}[x={(img.south east)}, y={(img.north west)}]

            \node[anchor=north west, xshift=0.02\linewidth, yshift=-0.02\linewidth]
                at (0.0,1.125) {\textsf{(a)}};

            \node[anchor=north east, xshift=-0.02\linewidth, yshift=-0.02\linewidth]
                at (0.49,1.125) {\textsf{(b)}};

        \end{scope}
    \end{tikzpicture}
    \caption{\textbf{Dependence of stabilization on frequency and amplitude.} \textbf{a)} Normalized surviving population following a 16-cycle pulse with a $\text{sin}^2$ envelope following the same procedure as in Fig \ref{fig:stabilization}(b), for drive frequencies of $0.89\,\omega_{\mathrm{0}}$ to $1.43\,\omega_{\mathrm{0}}$ (25 to 40~Hz). Dashed line indicates the point where the quiver amplitude equals $w/2$. Solid lines in corresponding shades represent the results of GPE numerics.  \textbf{b)} 2D stability diagram exploring the dependence of strong-field stabilization on frequency and amplitude over a wider range of amplitudes, in terms of maximum acceleration: $A\omega_{\text{drive}}^2$. Each pixel denotes the surviving bound population following a 16-cycle $\text{sin}^2$ pulse, normalized by the population in the undriven trap after the same duration. The data demonstrate that stabilization emerges at high amplitudes for all frequencies, with a frequency-dependent threshold amplitude.}
    \label{fig:PD}
\end{figure*}

\section{Results}
Our experimental investigation of strong-field stabilization focuses on this less-understood low-frequency regime. To probe the possible emergence of strong-field stabilization we first apply long flat-top pulses of varying quasistatic \textcolor{black}{amplitudes} \textcolor{black}{ at $\omega_{\mathrm{drive}}=1.43\omega_0$}, with a short 6-cycle ramp up from zero \textcolor{black}{amplitude} followed by an extended constant-\textcolor{black}{amplitude} drive. To measure the analogue of ionization, we suddenly terminate the pulse after some variable hold time by quenching both drive and trap to zero on timescales much faster than a drive cycle, and count the remaining atoms by absorption imaging after a brief period of free expansion. From these measurements, the ionization rate is extracted by fitting the remaining population as a function of pulse duration to a decaying exponential. Due to the need for fine control of pulse shape, such measurements would be considerably more challenging in pulsed laser experiments. 
The results of these measurements clearly demonstrate that the ionization rate does not grow monotonically with increasing amplitude, but rather increases to a maximum at intermediate amplitudes and then strongly decreases (Fig \ref{fig:stabilization}a). This constitutes a direct observation of strong-field stabilization.  

To probe strong-field stabilization under conditions more conducive to pulsed laser experiments, we next measure the surviving population following a complete non-truncated pulse with a more achievable shape, applying a 16-cycle pulse with a $\text{sin}^2$ envelope and a variable peak amplitude. Measurements of the surviving population fraction as a function of peak amplitude are shown in Fig~\ref{fig:stabilization}(b). 
While at lower amplitudes the surviving population shrinks as pulse amplitude grows, we observe  that the surviving population plateaus and recovers at higher amplitudes; this constitutes another demonstration of strong-field stabilization, now for a more directly experimentally-relevant pulse shape.
As a check on our understanding of the dynamics, Figure~\ref{fig:stabilization}(b) also includes numerical solutions of the one-dimensional Gross-Pitaevskii equation (GPE)~\cite{gross1961structure,pitaevskii1961vortex} which, while not capturing finite-temperature or beyond-mean-field effects, demonstrate qualitative agreement with experiments. 

To gain insight into the mechanism underlying strong-field stabilization, it is extremely helpful (but again challenging in pulsed-laser experiments) to be able to follow wavepacket evolution on timescales of individual optical cycles. 
To this end, we measured the dynamics of the driven atoms at successive drive cycles during a pulse, for pulse amplitudes both below and above the bifurcation threshold  ($\sim(8/9)w \approx 89.3\mu\text{m}$).
Results are depicted in \ref{fig:stabilization}(c) for quiver amplitudes below (red point in Fig. \ref{fig:stabilization}(b)) and above (blue point in Fig. \ref{fig:stabilization}(b)) the bifurcation threshold. We observe that densities follow the instantaneous potential below this threshold but accumulate near the turning points above it, tracking instead the cycle-averaged potential. \textcolor{black}{Below the bifurcation threshold, substantial loss occurs when the pulse approaches its peak; past this threshold unbinding from the trap occurs predominantly during the ramp-up and ramp-down of the pulse.} GPE simulations at similar amplitudes (Fig.~\ref{fig:stabilization}d) show behaviors similar to those seen experimentally: in the non‑stabilizing case, once the instantaneous acceleration becomes large enough, part of the wavepacket is emitted during each half‑cycle while the remainder continues to follow the shaken trap. In the stabilizing case, the density bifurcates and the two lobes remain near the turning points of the trap motion $a(t)$ with very small drift velocities, interacting with the trap only when it slows and producing a time‑averaged confinement through repeated emission and recapture. These complementary experimental and numerical results, intuitively surprising for drive frequencies so close to the trap frequency, illuminate the microscopic mechanism of strong-field stabilization and illustrate the central role of the cycle-averaged effective potential even in the low-frequency regime. 

Having established the clear observation of strong field stabilization and probed its mechanism, we can now experimentally map out the regimes of pulse parameters | most crucially frequency and amplitude | at which stabilization can occur.
We experimentally characterize the dependence of stabilization on pulse parameters by measuring the surviving population following a pulse of the same 16-cycle $\sin^2$ form for a wide range of frequencies and amplitudes. Fig \ref{fig:PD}a shows surviving population as a function of quiver amplitude for varying pulse frequencies. 
\textcolor{black}{For all driving frequencies, the measured loss increases with amplitude up to $A\approx w/2$, where the loss is stabilized and then ultimately reduced as the amplitude enters the stabilization regime. At such amplitudes, the condensate can no longer adiabatically track the moving trap \cite{SI}. While the reduction in loss is much more evident for driving frequencies greater than the trap frequency, even driving frequencies below the trap frequency clearly exhibit the plateau in the loss rate.  \textcolor{black}{The requisite amplitudes for stabilization in this regime surpass \textcolor{black}{the over-the-barrier intensity $I_{\mathrm{BSI}}(\mu)$}, where the drive-induced force exceeds the confining force of the trap, paralleling the predictions of Ref \cite{volkova_ionization_2011} (see also \cite{SI}).}
} 

\textcolor{black}{True stabilization in this lowest-frequency regime can be probed by extending the range of the pulse amplitude $A$ to larger values.  Because the inertial force, and therefore the effective electric field, scales with $\omega_\text{drive}^2$, the amplitude can be scaled to keep the peak force the same across all frequencies. Taking this approach, we measured a 2D stability diagram as a function of frequency and maximum acceleration (linearly proportional to effective electric field). The results are shown in Fig~\ref{fig:PD}(b). The corresponding amplitudes exceed the maximum amplitude presented in Fig \ref{fig:PD}(a) by up to a factor of 4. 
We observe that suppression of loss occurs even for frequencies below the trap frequency, for which loss appeared to plateau over the smaller range of amplitudes given in Fig. \ref{fig:PD}(a): lower frequencies require higher applied forces to induce stabilization. \textcolor{black}{On the parameter map of Fig.~\ref{fig:params}, these drives ($K_0\gtrsim200$) sit deep in the low-frequency strong-field limit, far from the high-frequency limit in which stabilization was originally predicted; the measurements thus directly explore this previously untested regime.}
For each drive frequency we observe a region of maximum loss when the amplitude approaches approximately $w/2$. The dark-colored swath of the stability diagram thus represents the first experimental observation of the ``death valley" phenomenon often discussed in the strong-field stabilization literature \cite{faria_influence_1998, gavrila_atomic_2002}. Time-resolved data like those shown in Fig.~\ref{fig:stabilization} indicate that in high-amplitude pulses, the majority of the loss occurs as the pulse ramps up through this regime of minimum stability. }

\begin{figure}[ht!]
    \centering
    \includegraphics[width=0.9\columnwidth]{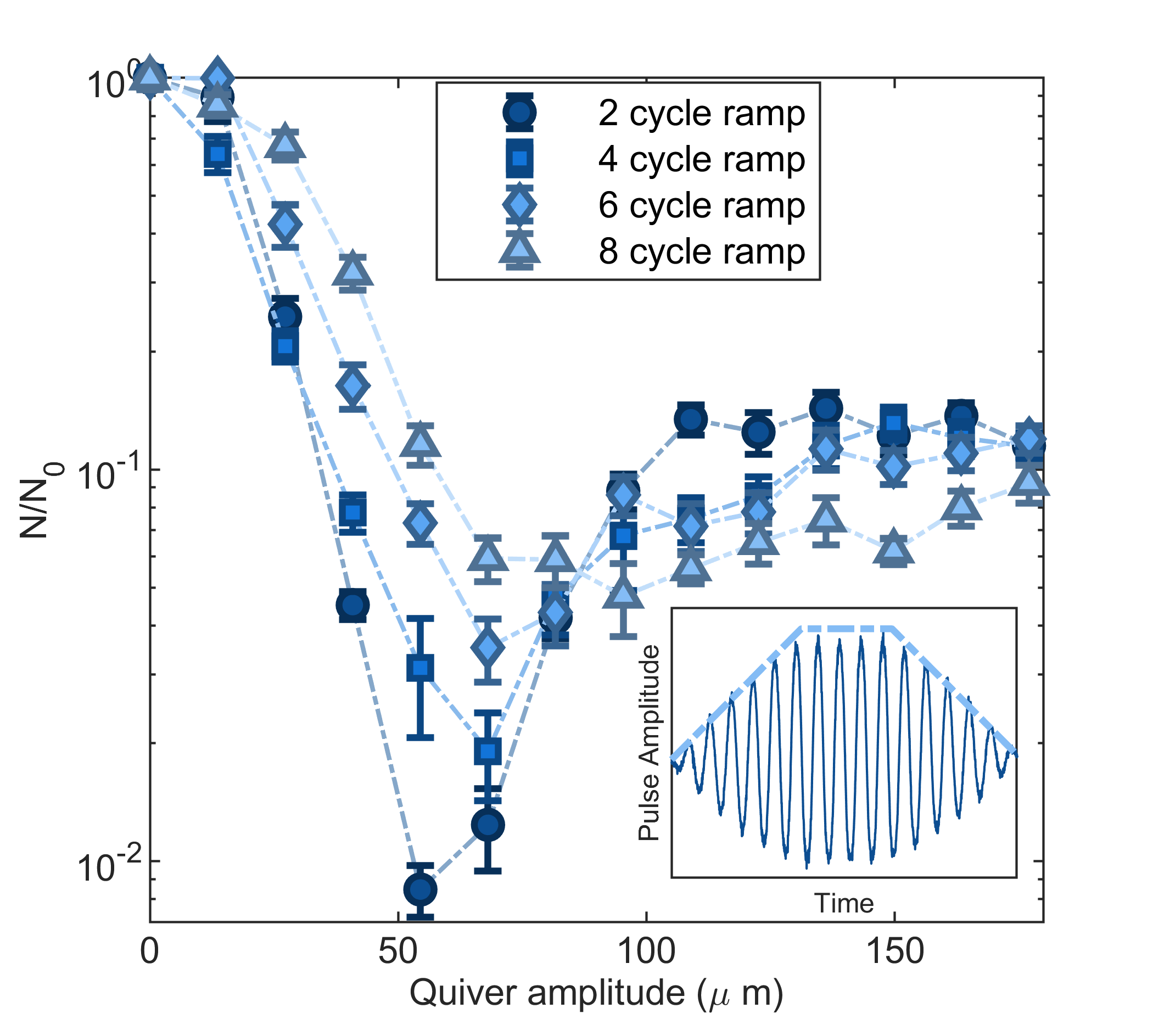}
    \caption{\textbf{Effect of pulse shape on ionization} Measured surviving population versus amplitude following a 16-cycle trapezoidal pulse of frequency $1.34\,\omega_{\mathrm{0}}$ (37.5~Hz), with the duration of the ramp-up and ramp-down varied. Error bars represent standard error; lines are guides to the eye. Inset shows the trap center motion during a 6-cycle ramp, as measured by a position-sensitive photodiode.}
    \label{fig:Panel}
\end{figure}

The idea  of such a ``death valley" as a practical barrier to the observation of strong field stabilization leads naturally to a consideration of the detailed effects of the pulse shape in a realistic laser experiment~\cite{gavrila_atomic_2002, fedorov_stabilization_1999}. We apply the fine control of pulse envelope available in our experimental platform to a direct experimental investigation of pulse-shape-dependent phenomena, 
with the results shown in Fig.~\ref{fig:Panel}. We observe that for a trapezoidal pulse of fixed total duration, the amplitude-dependent ionization curves are strongly modified by changing the ramp durations.  
Shorter ramps lead to enhanced loss below the stabilization threshold, as moderate amplitudes for which the decay rate is largest occupy a  greater duration of the pulse. This trend reverses for peak amplitudes surpassing the stabilization threshold, where it is favorable to quickly ramp through the ``death valley'' to the stabilization regime. These experimental results quantitatively illuminate the important role of the pulse envelope in shaping unbinding dynamics in strong fields. 

\section{Discussion}
Using a strongly driven trapped-atom platform to emulate the dynamics of bound electrons in a strong laser pulse, we have demonstrated strong-field stabilization of a quantum mechanical ground state and characterized the resulting dynamics across a swath of parameter space including the regime of drive frequencies below the natural frequency of the binding potential. The results affirm that stabilization persists well beyond the high-frequency regime. Experimental and numerical measurements with different atom numbers~\cite{SI} demonstrate  stabilization  across this full range, confirming that the effect does not depend on mean-field interactions and reflects a generic property of strongly driven bound systems.
We have imaged the spatial bifurcation of the wavepacket and directly observed the structural changes associated with the Kramers-Henneberger potential. We have used the fine control available in this platform to quantitatively explore the effects of pulse shape on ionization, showcasing the ability of this novel experimental approach to investigate practical issues of strong-field dynamics relevant to other (e.g. pulsed-laser) platforms.

The experimental platform presented here is well-suited to the study of strong-field quantum dynamics mainly due to three features: accessible timescales, a high degree of controllability, and the capacity for time-resolved measurement of multiple observables, free of the focal-volume averaging over many atoms that can obscure the single-atom response. These capabilities open up numerous possible directions for future work.
One example would be direct table-top emulation of the non-perturbative, high-intensity regime that future free-electron-laser
facilities aim to reach: near-future capabilities of X-ray free-electron laser facilities~\cite{montironi2019xleap,duris2020tunable,walker2024above} motivate using this platform to experimentally explore high-frequency stabilization~\cite{boitsov2026quasiperiodicdynamicsnondipolexray} and the collective response of heavy atoms in intense X-ray fields~\cite{ball1973photoabsorption}. 
Other natural extensions include probing coherent control of strong-field ionization with elliptical polarization~\cite{reiss_frequency_1992,Olofsson_Fulton_Tahouri_Bertolino_Djiokap_Dahlstrom_2026} as well as testing the limits of the Born-Oppenheimer approximation and various effective potentials. More broadly, the approach presented here represents an opportunity for closer collaboration on quantum dynamics between the largely disjoint ultrafast and ultracold communities of atomic physics. 

\section{Methods summary}
\subsection{Experimental details}
The moving optical trap is formed by a pair of crossed focused beams of 1064 nm light; one beam is translated while the other remains static. The intensity of these beams can be independently controlled. The translatable or ``shaken'' beam passes through an acousto-optic modulator (AOM); the first deflected order is used for the trap and is focused by a lens with focal length $f_{\text{lens}}=150 \text{ mm}$ and aligned to overlap with the focus of the static beam. The beam waist $w$ is $100.5 \,\mu\text{m}$ and the typical trap depth is $V_0=7.82\pm0.30~\mathrm{E_{R}}$, where the recoil energy $E_{\mathrm{R}} = \hbar^2k^2/2m$ with $k=2\pi/\lambda_{\text{trap}}$. This setup is depicted in Fig 1(a). The trap center is translated by modulating the frequency of the radiofrequency (RF) drive of the AOM, 
thereby shifting the deflection angle of the shaken beam,  which depends on RF frequency $\nu_\mathrm{RF}$ as  $\theta_\mathrm{AOM}=(\lambda_\mathrm{trap} \times \nu_\mathrm{RF})/v_\mathrm{AOM}$, where $v_\mathrm{AOM}$ is the speed of sound in the AOM crystal. Modulation of $f_\mathrm{RF}$ leads to small changes in $\theta_\mathrm{AOM}$. The effect at the focus is a lateral shift of $d=f_\mathrm{lens}\Delta\theta_\mathrm{AOM}$, along the direction of the orthogonally propagating static beam. 

\subsection{Calibration of Experimental Parameters \label{sec:calibration}}
Measuring the trap depth $V_0$ of the static trap requires measurement of the trap frequency as well as accurate knowledge of the beam waist. For a static Gaussian trap, the trap  frequency $\omega_0 = 2\pi f_0$ is given by $V_0 = m (\pi w f_0)^2$ where $m$ is the mass of the strontium atom. The harmonic frequency $f_0$ is measured by inducing a center-of-mass oscillation of the condensate using the translatable beam to be $27.98 \pm 0.28~\text{Hz}$, where the error reported represents the $95\%$ confidence interval. Together with the value of the beam waist $w$, measured to be $100.5\pm1.5~\mu\text{m}$ the trap depth is directly calculated to be $7.82\pm0.30~\mathrm{E_{R}}$. 

The beam waist and other experimental parameters can be calibrated and cross-checked using the properties of the deformed trap itself. As discussed in the main text and elaborated on in \cite{SI}, the time-averaged potential of a periodically translated single-well potential demonstrates dichotomy: above a critical amplitude the time-averaged trap deforms and bifurcates into a double-well potential. For a Gaussian trap, the form of the time-averaged potential and the critical amplitude can be analytically evaluated. These analytical results can be also be used to calibrate the relevant experimental parameters. Here we only include the relevant specific results; further detailed discussion of the time-averaged potential and its properties may be found in \cite{SI}. We compute analytically the equation that dimensionless local potential extrema $z_\pm := x_\pm/w$ satisfy from the time-averaged potential as a function of the dimensionless drive amplitude $\beta$. Together with the critical bifurcation amplitude $\beta_c$, which may be determined analytically from the form of the time-averaged potential \cite{SI}, the dimensional, spatial drive amplitude $A$ can be related to the experimentally controlled analog voltage $V_\mathrm{drive}$
which sets the frequency modulation and to the beam waist $w$. At large drive frequencies, only the time-averaged term contributes and the \textit{in situ} condensate positions directly reflect the trap minima at different drive amplitudes $A$ and cleanly demonstrate dichotomy as shown in Fig. 1(c). These measurements can then provide experimental characterizations of the deformed trap, and the beam waist can be cross-checked with the experimentally measured locations of the bifurcated trap minima. 

The procedure for this measurement is as follows: the position modulation of the trap center is ramped up with a drive frequency of $5\,\mathrm{kHz}$, orders of magnitude larger than the harmonic frequency of the static trap so that the time-averaged description holds. 
The condensates are then measured \textit{in situ} and their centers of mass are extracted by a Gaussian fit. We subtract all the fitted centers, taking only one branch, denoted as $x_+(V_\mathrm{drive})$, and implicitly fit them to $p_1 z_+(p_2) = x_+(V_\mathrm{drive})$.
The fit extracts the spatial unit and the critical bifurcation amplitude by solving for two parameters: $p_1:= x_+/z_+ = w$ converts the dimensionless spatial coordinate to the physical unit of micrometers, and thus measures the beam waist $w$; $p_2:= \beta/V_\mathrm{drive}$ converts the shaking amplitude in volts $V_\mathrm{drive}$ to the dimensionless shaking amplitude $\beta$. The fitted critical amplitude $v_c$ is related to the dimensionless one $\beta_c$ as $v_c = \beta_c/p_2$.
With the knowledge of the beam waist $w$ (or equivalently $p_1$), the spatial shaking amplitude $A$ is then given by $A = p_1 p_2 V_\mathrm{drive}$, thus calibrating the relationship between the voltage $V_{\mathrm{drive}}$ and the spatial amplitude $A$. The dichotomy dataset, exemplified by Figure 1(c) in the main text, contains all the information to calibrate beam parameters.

\subsection{Numerical modeling}
The dynamics of the BEC in the shaken trap are modeled using the one-dimensional GPE,
\begin{equation}\label{eq:GPE}
\left[\text{i}\hbar\,\partial_{t}
      +\frac{\hbar^{2}}{2m}\partial_{x}^{2}
      -V_{\text{trap}}
      -gN_{0}|\psi|^{2}
      +\frac{\text{i}}{2}\Gamma(x)
\right]\psi(x,t)=0,
\end{equation}
which we solve numerically on a finite spatial grid.  
The effective 1D interaction strength is 
$g \approx 4\pi\hbar^{2}a_{\text{s}}/(m a_{\text{ho}}^{2})$~\cite{dum_wave_1998}, where 
$a_{\text{s}} \approx 123\,a_\text{B} \approx 3.14\times 10^{-3}\,a_{\text{ho}}$ is the scattering length,
$a_\text{B}$ is the hydrogenic Bohr radius, and 
$a_{\text{ho}}=\sqrt{\hbar/(m\omega_0)}$ is the harmonic-oscillator length.  
A complex absorbing potential $-\text{i}\Gamma(x)/2$ imposes outgoing boundary conditions for atoms
leaving both the trap and the numerical domain.  
Further numerical details, together with a full analysis of the stabilization mechanism and an extended discussion of how stabilization emerges across different driving regimes—including qualitative differences between single‑atom behavior and the collisional limits of $N_0=4\times10^{4}$, $1\times10^{5}$, and $2.5\times10^{5}$ initially bound atoms—are presented in the Supplementary Information~\cite{SI}.
\subsection{Data availability}
Data is available from the corresponding author upon request. 
\section{End notes}
\subsection{Acknowledgements}
We acknowledge Shravan Ramanand for experimental assistance and a critical reading of the manuscript, Xiao Chai for providing the MuscleMuseum control system (manuscript in preparation), and Jeremy Tanlimco for a critical reading of the manuscript. ASL and SW acknowledge U.S. Department of Energy, Office of Basic Energy Sciences, Atomic, Molecular, and Optical Sciences Program, Award ID DESC0022093. ARD, YB, PK, and DMW acknowledge support from the NSF QLCI program (OMA-2016245), the Noyce Foundation, the Eddleman Quantum Institute, and the UC Santa Barbara NSF Quantum Foundry funded via the Q-AMASE-i program under Grant No. DMR1906325. ARD acknowledges support from the NSF NRT program under Grant No. 2152201. 

\subsection{Author Information}

ARD, YB, and PK prepared the experimental setup and performed the experiments and data analysis.  SW and ASL performed theoretical calculations. DMW supervised the project. All authors contributed substantially to the work presented in this paper, including discussions of data and preparation of the manuscript. Correspondence and requests for materials should be directed to DMW.

\subsection{Ethical declarations}
The authors declare no competing interests.
\subsection{Additional Information}
Supplemental information is available for this text.
\bibliography{Bib1}
\end{document}


\section*{Supplemental Information}
\setcounter{section}{0}
\textcolor{black}{Stabilization generally occurs when the drive-induced length and velocity scales of the bound system exceed the natural dynamical scales of the undriven system. While the microscopic details are system-dependent, these generic criteria for stabilization arise in varying systems, from strong-field atomic physics to the many-body response of an interacting condensate.}
This  appendix provides the quantitative framework used to interpret the dynamics observed
in our experiment. It is organized to parallel the elements needed to understand
the measurements: Sec. \ref{sec:GPE} introduces the Gross--Pitaevskii equation; Sec. \ref{sec:trap} describes the
shaken trap potential; Sec. \ref{sec:SO_FFT} presents the numerical method and the remaining-fraction data
used throughout the paper; Sec. \ref{sec:FFT_appendix_hydro} develops a hydrodynamic description \textcolor{black}{which outlines the natural dynamical scales governing the condensate response to the driven trap}; Sec. \ref{sec:FFT_appendix_bohm} analyzes
microscopic dynamics using Bohmian trajectories; Sec. \ref{sec:slow_driving} examines stabilization in the
slow-driving regime; Sec. \ref{sec:one_atom} \textcolor{black}{discusses the role of atom number, presenting experimental data at varied atom number, and} models stabilization of a single trapped atom in the absence of $s$-wave
scattering, highlighting the fact that interactions in the condensate produce a qualitatively different form of
stabilization with distinct characteristic scales; and Sec. \ref{sec:strong-field_emulation} discusses the broader context
of driven stabilization and its connections to strong-field physics.

The simulations used throughout this appendix are based on the Gross--Pitaevskii equation \cite{gross1961structure,pitaevskii1961vortex}
introduced in Sec. \ref{sec:GPE} and are evolved numerically using the fourth-order split-operator
method of Yoshida \cite{yoshida1990construction} described in Sec. \ref{sec:SO_FFT}. Throughout the appendix we interpret lengths,
energies, and times in terms of the harmonic-oscillator scales
$a_\text{HO}=\sqrt{\hbar/(m\omega_0)}=2.07\,\mu m$,
$\hbar\omega_0$, and $\omega_0^{-1}$, where
$\omega_0=2\pi f_0$ with $f_0=27.9861$ Hz and atomic
mass $m\approx 84$ amu for ${}^{84}\mathrm{Sr}$. In the numerical implementation we set
$\hbar=m=\omega_0=a_\text{HO}=1$, and the Yoshida split-operator scheme in
Sec. \ref{sec:SO_FFT} is written in this dimensionless form.
Our simulations extend the parameter space explored by Dum \textit{et al.}\
\cite{dum_wave_1998}, who studied shaken BEC traps with
$a_0\in[10\text{--}50]\,a_\text{HO}$, $gN_0\in[0\text{--}100]$,
$\omega_{\text{drive}}/\omega_0\in[2\text{--}20]$, and evolution over hundreds of
cycles. We explore different regimes: (1) larger amplitudes, with
$a_0\in[0\text{--}180]\,\mu m$ ($[0\text{--}87]\,a_\text{HO}$, up to
$1.7\times$ larger) for standard sweeps and $a_0\in[200\text{--}700]\,\mu m$
($[97\text{--}338]\,a_\text{HO}$, up to $6.8\times$ larger) for the extended 25 Hz sweep;
(2) much stronger interactions, with $gN_0\approx1578$ for $N_0=40{,}000$,
$gN_0\approx3945$ for $N_0=100{,}000$, and $gN_0\approx9862$ for $N_0=250{,}000$
(16$\times$ to 100$\times$ larger than Dum \textit{et al.}); (3) near-resonant frequencies,
with $f_{\text{drive}}\in\{25,30,35,40\}$ Hz giving
$\omega_{\text{drive}}/\omega_0\in[0.89\text{--}1.43]$ rather than the
higher-frequency regime of Dum \textit{et al.}; and (4) shorter evolution times consisting of
16 cycles of driven motion followed by 16 cycles of free propagation to remove untrapped
atoms from the computational domain. Additionally, we perform amplitude sweeps at
$f_{\text{drive}}=40$ Hz ($\omega_{\text{drive}}/\omega_0=1.43$) with
$N_0=40{,}000$ and $N_0=100{,}000$ atoms using identical simulation parameters to
investigate the role of interaction strength, and we also compute single-atom ($N_0=1$)
results at the same 40 Hz drive with $s$-wave scattering removed to provide a direct
comparison with the non-interacting limit.

\section{Gross-Pitaevskii Equation}\label{sec:GPE}

The condensate wavefunction $\psi(x,t)$ evolves according to \cite{pitaevskii2016bose}
\begin{equation}
\text{i}\hbar\frac{\partial}{\partial t}\psi(x,t)
=
\left[
- \frac{\hbar^2}{2m}\frac{\partial^2}{\partial x^2}
+ V_{\text{trap}}(x,t)
+ g(N_0-1)|\psi(x,t)|^2
- \frac{\text{i}}{2}\Gamma(x)
\right]\psi(x,t),
\end{equation}
where the nonlinear term represents the mean-field potential generated by low-energy
$s$‑wave collisions between atoms in the condensate. These collisions dominate elastic
interactions in ultracold gases and give rise to an effective one-dimensional coupling
strength $g \approx 4\pi\hbar^2 a_s/(m a_\text{HO}^2)$ \cite{dum_wave_1998}, with
$a_s \approx 123\,a_\text{B} \approx 3.14\times10^{-3}\,a_\text{HO}$ the $s$‑wave
scattering length. The factor $(N_0-1)$ accounts for the number of interacting partners per
atom in the mean-field limit.

The shaken trap potential is implemented as
\begin{equation}
V_{\text{trap}}(x,t)
=
- V_0 \exp\!\left[-\frac{2(x-a(t))^2}{w^2}\right],
\end{equation}
with trap depth $V_0 = 0.25\,m\omega_0^2 w^2$ and waist
$w = 100.5\,\mu m$. The trap center follows
\begin{equation}
a(t)=
\begin{cases}
a_0(t)\sin(\omega_{\text{drive}} t), & t < t_{\text{pulse}}, \\
0, & t \ge t_{\text{pulse}},
\end{cases}
\end{equation}
where $\omega_{\text{drive}} = 2\pi f_{\text{drive}}$ and the envelope
$a_0(t)=a_0\sin^2(\pi t/t_{\text{pulse}})$ ramps the drive smoothly on and off. The pulse
duration
\begin{equation}
t_{\text{pulse}} = 16 \times \frac{2\pi}{\omega_{\text{drive}}},
\end{equation}
corresponds to 16 drive cycles, followed by 16 cycles of free evolution with the trap held
stationary ($a=0$).

Open boundary conditions are implemented using the complex absorbing potential
$-\text{i}\Gamma(x)/2$ \cite{riss1993calculation}:
\begin{equation}
\Gamma(x)=
\begin{cases}
20\hbar\omega_0
\left(\dfrac{|x|-0.9L_x/2}{0.1L_x/2}\right)^2, & |x|>0.9L_x/2,\\
0, & \text{otherwise}.
\end{cases}
\end{equation}

With the GPE and interaction parameters established, we now examine the structure of the
shaken trap potential, beginning with its unshaken limit.

\section{Properties of the Shaken Trap Potential}\label{sec:trap}

When the shaking amplitude goes to zero ($\beta \equiv a_0/w \to 0$), the Gaussian trap
reduces near the center ($|x| \ll w$) to a harmonic oscillator,
\begin{equation}
    V_\text{trap}(x) \xrightarrow{|x|\ll w} V_\text{trap}(0) + \frac{1}{2}m\omega_0^2 x^2,
\end{equation}
with trap depth $V_\text{trap}(0) = -V_0$ and unshaken (angular) oscillator frequency
\begin{equation}
    \omega_0 = 2\pi f_0 = \frac{2}{w}\sqrt{\frac{V_0}{m}},
\end{equation}
obtained by matching $\frac{1}{2}m\omega_0^2 x^2 = \frac{2V_0}{w^2} x^2$.

Deeply bound single-particle levels therefore exhibit nearly harmonic behavior, with
approximate energies
\begin{equation}
    \varepsilon_n \approx -V_0 + \hbar\omega_0\left(n + \tfrac{1}{2}\right).
\end{equation}
At higher energies, however, the Gaussian deviates from a parabola, and the level spacing
departs from the uniform harmonic-oscillator pattern.

\begin{figure}[t]
\centering
\includegraphics[width=0.8\textwidth]{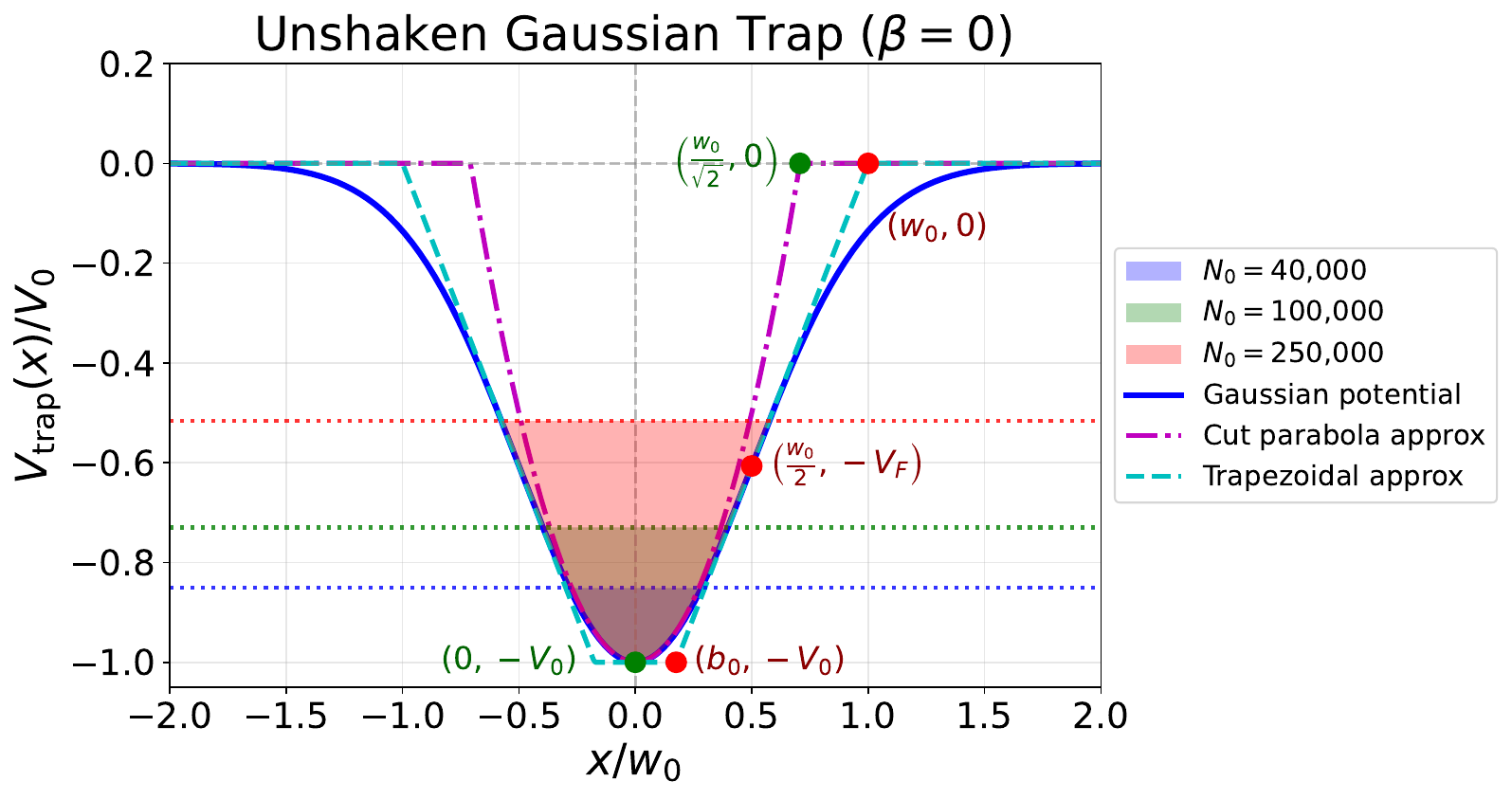}
\caption{Unshaken Gaussian trap potential ($\beta = 0$) as a function of position scaled by the
beam waist $w$. Blue solid: Gaussian potential 
$V_{\mathrm{trap}}(x) = -V_0 \exp(-2x^2/w^2)$. 
Magenta dash--dot: cut--parabola approximation (harmonic oscillator truncated at $V=0$). 
Cyan dashed: trapezoidal approximation. The potential depth is $V_0$. Horizontal lines denote 
the chemical potentials for condensates with $N_0 = 40\,$k, $100\,$k, and $250\,$k, and the 
corresponding shaded regions indicate the Thomas--Fermi fillings for each case.}
\label{fig:unshaken}
\end{figure}

Figure~\ref{fig:unshaken} shows the unshaken Gaussian trap ($\beta = 0$). The trap has depth
$V_0$ at its center and the usual Gaussian profile set by the beam waist $w$. The
inflection points at $x=\pm w/2$ mark the locations of maximum restoring force. For
$|x|\ll w$, the potential is well approximated by a harmonic oscillator with frequency
$\omega_0$. The magenta dash–dot curve shows the “cut parabola’’ approximation,
$V(x) = -V_0 + \tfrac{1}{2}m\omega_0^2 x^2$, truncated at $V_\text{trap}=0$ at
$|x| = w/\sqrt{2}$, similar to the model used by Dum \textit{et al.}~\cite{dum_wave_1998}.
This approximation, however, reaches $V_\text{trap}=0$ too quickly and significantly overestimates the restoring force at outer edges of the trap potential. For this
reason, our numerical GPE solutions use the full Gaussian form.

A trapezoidal approximation (cyan dashed curve in Fig.~\ref{fig:unshaken}) provides a more
accurate global analytic model for the unshaken trap. For the unshaken case,
\begin{equation}
    V(x)=
    \begin{cases}
        -V_0, & |x|<b_0, \\
        -V_0 + F(|x|-b_0) = -V_F + F(|x| - b_F), & b_0 \le |x| < b_1, \\
        0, & |x|\ge b_1,
    \end{cases}
\end{equation}
where the linear sections are tangent to the Gaussian at its point of maximum force,
$F = (2V_0/w)e^{-1/2}$, occurring at $|x|= b_F \equiv w/2$ with corresponding potential
value $V_F = V_0 e^{-1/2}$.

The flat–to–ramp boundary lies at $x=\pm b_0$, where
\begin{equation}
    b_0 = b_F - \frac{V_0 - V_F}{F}
        = \frac{w}{2} - \left(1-e^{-1/2}\right)\frac{V_0}{F}.
\end{equation}
The ramp terminates at $x=\pm b_1$, where the potential is truncated to zero:
\begin{equation}
    b_1 = b_0 + \frac{V_0}{F}
        = b_F + \frac{V_F}{F}
        = w.
\end{equation}
The resulting trapezoidal potential is shown as the cyan dashed curve in
Fig.~\ref{fig:unshaken}.

In the stabilization regime, both experiment and simulation show that the condensate density
closely follows the structure of the cycle-averaged Kramers--Henneberger \cite{kramers1956collected,henneberger1968perturbation} potential
\begin{equation}
    V_\text{avg}(x)
    \equiv \frac{1}{2\pi}\int_{-\pi}^{\pi} d\varphi\,
    V_\text{trap}\!\left(x - a_0 \sin\varphi\right),
\end{equation}
where $V_\text{trap}(x) = -V_0 e^{-2x^2/w^2}$ and $\beta \equiv a_0/w$. Using
\begin{equation}
    V_\text{trap}\!\left(x - a_0 \sin\varphi\right)
    = -V_0 \exp\!\left[-\frac{2}{w^2}\left(x - a_0 \sin\varphi\right)^2\right],
\end{equation}
the exponent may be rearranged to give
\begin{equation}
    V_\text{avg}(x)
    = -V_0 e^{-\beta^2} e^{-2x^2/w^2}
      \frac{1}{2\pi}\int_{-\pi}^{\pi} d\varphi\,
      \exp\!\left[4\beta\frac{x}{w}\sin\varphi
      + \beta^2\cos(2\varphi)\right].
\end{equation}

Using the Jacobi--Anger expansions \cite{DLMF10.35}
\begin{equation}
    e^{z\cos\varphi} = \sum_{n=-\infty}^{\infty} I_n(z)\,e^{\text{i} n\varphi},\qquad
    e^{z\sin\varphi} = \sum_{n=-\infty}^{\infty} (-\text{i})^n I_n(z)\,e^{\text{i} n\varphi},
\end{equation}
the integral evaluates to the Bessel-series representation
\begin{equation}
    V_\text{avg}(x)
    = e^{-\beta^2} V_\text{trap}(x)
      \sum_{n=-\infty}^{\infty} (-1)^n
      I_{2n}\!\left(4\beta\frac{x}{w}\right) I_n(\beta^2), \label{eq:Vavg}
\end{equation}
which is strictly even in $x$. We consider the average over a single shaking cycle, assuming that the time-dependent envelope \(a_0(t)\) varies slowly enough during the cycle that it can be treated adiabatically as a constant \(a_0\). In later sections, we show that this approximation is sufficient for interpreting our results. However, non-adiabatic effects may become important when shaking for only a few cycles, rather than the 16 used here.

To determine when the trap center changes stability, we expand $V_\text{avg}(x)$ near $x=0$.
Using the small-argument expansions of the modified Bessel functions and retaining only the
terms contributing to the curvature yields
\begin{equation}
    V_\text{avg}''(0)
    = \frac{4V_0}{w^2} e^{-\beta^2}
      \left[(1 - 2\beta^2) I_0(\beta^2)
      + 2\beta^2 I_1(\beta^2)\right].
\end{equation}
The critical shaking amplitude $\beta_c$ is defined by $V_\text{avg}''(0)=0$, i.e.
\begin{equation}
    (1 - 2\beta_c^2) I_0(\beta_c^2)
    + 2\beta_c^2 I_1(\beta_c^2) = 0,
\end{equation}
or equivalently
\begin{equation}
    \frac{I_1(\beta_c^2)}{I_0(\beta_c^2)}
    = 1 - \frac{1}{2\beta_c^2}.
\end{equation}
Solving gives
\begin{equation}
    \beta_c = \frac{8}{9} - \epsilon,
    \qquad
    \epsilon \approx 8.16\times 10^{-5}. \label{eq:criticalBeta}
\end{equation}

\begin{figure}[H]
\centering
\includegraphics[width=0.60\textwidth]{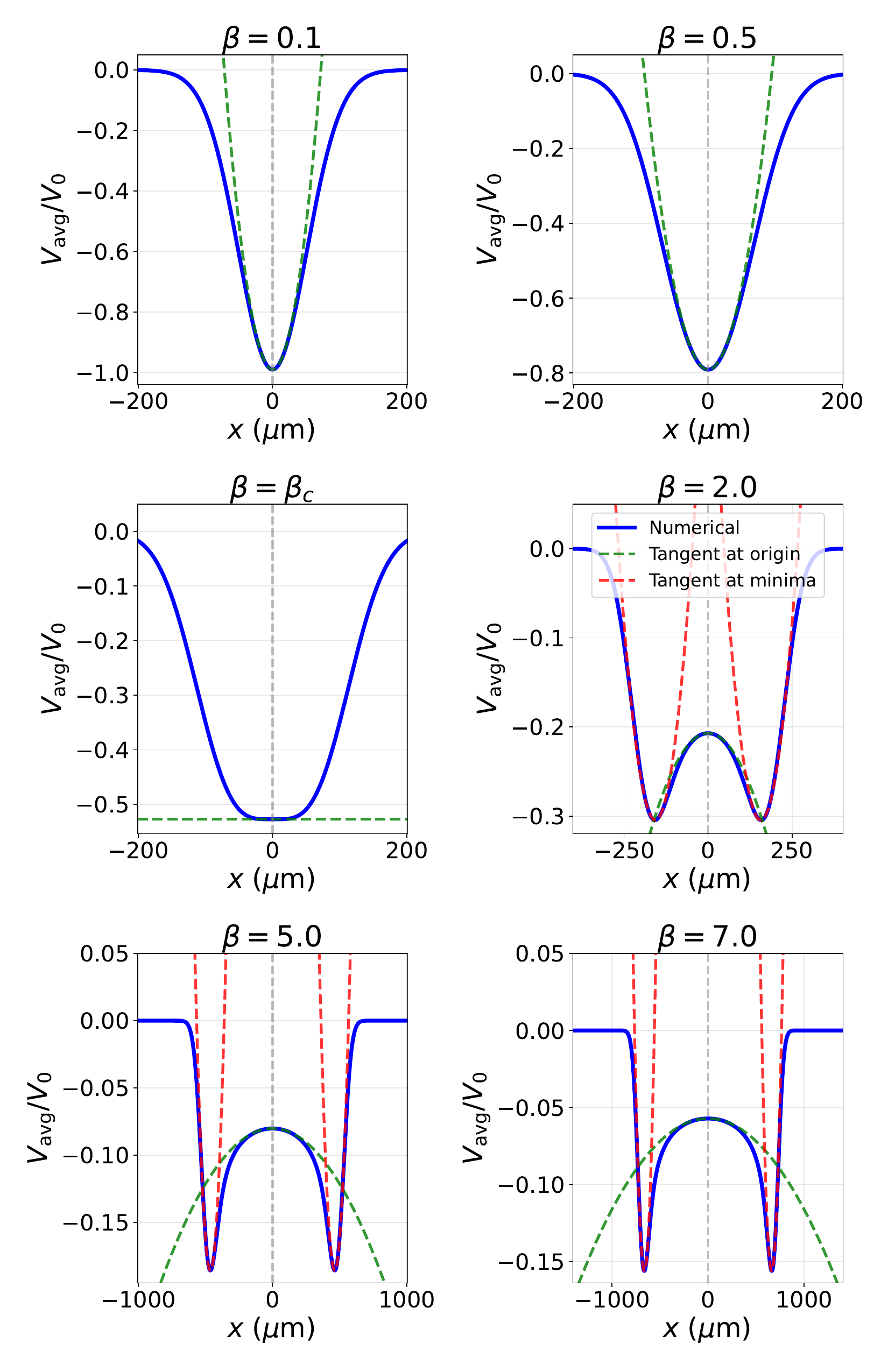}
\caption{Cycle-averaged potential $V_\text{avg}(x)/V_0$ for six values of $\beta$ with beam waist $w = 100.5~\mu$m. Blue solid curves: numerical potential. Green dashed curves: tangent parabolas at the origin. Red dashed curves: tangent parabolas at the bifurcated minima. The transition from a single-well structure ($\beta < \beta_c$) to a double-well structure ($\beta > \beta_c$) is evident.}
\label{fig:potentials}
\end{figure}

For $\beta < \beta_c$, the time-averaged potential has a single minimum at the origin. For
$\beta > \beta_c$, the curvature at $x=0$ becomes negative and the potential bifurcates into
a symmetric double well. Figure~\ref{fig:potentials} shows $V_\text{avg}(x)/V_0$ for several
values of $\beta$ across this transition. For weak shaking ($\beta = 0.1$), the averaged
potential remains close to the unshaken trap. As $\beta$ increases to 0.5, the trap widens
while retaining its qualitative shape. Near the critical value $\beta_c \approx 8/9 -
\epsilon$, the curvature at the origin approaches zero, as indicated by the tangent
parabolas marking local extrema. For $\beta > \beta_c$, a central barrier forms and the
potential develops a double-well structure. As $\beta$ increases further, the potential
flattens near $x=0$ and the bifurcated minima move outward toward $|x| \approx a_0$. This
reflects the fact that the shaken trap spends more time near its classical turning points at
$x = \pm a_0$, where its velocity is smallest, and moves most rapidly near the origin.
Consequently, atoms tend to accumulate where the trap resides most often.

\section{Split-Operator Method}\label{sec:SO_FFT}

\begin{figure}[t]
    \centering
    \includegraphics[width=\linewidth,
        trim=0cm 8.8cm 0cm 0cm, clip]{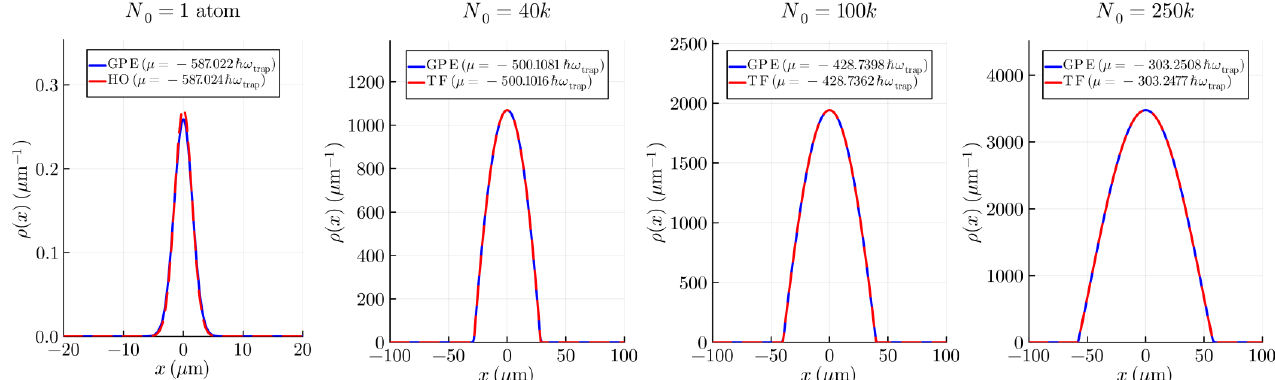}
    \caption{Ground-state density profiles obtained from the GPE (solid blue) and from the harmonic-oscillator (HO) approximation for $N_0 = 1$ (dashed red), together with the Thomas--Fermi (TF) approximation for $N_0 = 40{,}000$, $100{,}000$, and $250{,}000$ (also dashed red). The corresponding chemical potentials $\mu$ for each case are listed in the legend in trap units of $\hbar\omega_\text{trap}=\hbar \omega_0$.}
    \label{fig:GPE_HO_TF_ground_state}
\end{figure}

The Hamiltonian decomposes as $\hat{H} = \hat{T} + \hat{V}$, where
$\hat{T} = -\tfrac{1}{2}\,\partial^2/\partial x^2$ and
\begin{equation}
\hat{V} = V_{\text{trap}} + g(N_0 - 1)\,|\psi|^2 - \tfrac{\mathrm{i}}{2}\Gamma .
\end{equation}
(in dimensionless units). Because the GPE is defined with the prefactor
$g(N_0 - 1)$, we approximate it as $gN_0$ in simulations for large atom
numbers ($40$k, $100$k, and $250$k), since $N_0 \gg 1$ and the difference
between $g(N_0 - 1)$ and $gN_0$ is negligible for the behavior we study.
For the single-atom dataset, however, we retain the exact prefactor
$g(N_0 - 1)$, ensuring that the interaction term vanishes when $N_0 = 1$.
This makes the single-atom case strictly non-interacting while leaving the
multi-atom simulations effectively unchanged.

Potential operators $e^{-\mathrm{i}\hat{V}\Delta t}$ are applied via pointwise multiplication in
position space, while kinetic operators $e^{-\mathrm{i}\hat{T}\Delta t}$ are diagonal in momentum
space:
\begin{equation}
\psi(x) \xrightarrow[\mathcal{F}]{\mathrm{FFT}} \tilde{\psi}(k)
\xrightarrow{\times e^{-\mathrm{i}k^2\Delta t/2}} \tilde{\psi}'(k)
\xrightarrow[\mathcal{F}^{-1}]{\mathrm{IFFT}} \psi'(x).
\end{equation}

Real-time evolution uses Yoshida's fourth-order symplectic integrator
\cite{yoshida1990construction}:
\begin{equation}
\hat{U}_4(\Delta t)
= e^{-\mathrm{i}\hat{V}c_4\Delta t}
  e^{-\mathrm{i}\hat{T}d_3\Delta t}
  e^{-\mathrm{i}\hat{V}c_3\Delta t}
  e^{-\mathrm{i}\hat{T}d_2\Delta t}
  e^{-\mathrm{i}\hat{V}c_2\Delta t}
  e^{-\mathrm{i}\hat{T}d_1\Delta t}
  e^{-\mathrm{i}\hat{V}c_1\Delta t}
  + O(\Delta t^5).
\end{equation}

with coefficients
\begin{align}
\chi_0 &= -2^{1/3}/(2 - 2^{1/3}), \qquad
\chi_1 = 1/(2 - 2^{1/3}), \\
c_1 = c_4 &= \chi_1/2, \qquad
c_2 = c_3 = (\chi_0 + \chi_1)/2, \\
d_1 = d_3 &= \chi_1, \qquad
d_2 = \chi_0.
\end{align}

This scheme provides global accuracy of $O(\Delta t^4)$. Real-time evolution is used to compute the system dynamics, while imaginary-time propagation—implemented by replacing $\Delta t \rightarrow -\mathrm{i}\Delta\tau$ and renormalizing after each step—is used to obtain the initial state.

The spatial domain extends from $-L_x/2$ to $L_x/2$ and is discretized with $N_x$ uniformly spaced grid points, giving grid spacing $\Delta x = L_x/N_x$. The timestep $\Delta t$ is chosen to ensure convergence of the fourth-order integrator. During the evolution we save a fixed number of snapshots at 2,000 regular intervals for subsequent analysis. All numerical values are listed in Table~\ref{tab:parameters}.

\begin{center}
\begin{tabular}{@{}ll@{}}
\toprule
\textbf{Parameter} & \textbf{Value} \\
\midrule
Harmonic oscillator length & $a_\text{HO} = 2.07 \mu m$ \\
Unperturbed trap frequency & $f_{\text{HO}} = 27.9861$ Hz \\
Spatial domain & $x \in [-350, 350]\,a_\text{HO} = [-724, 724] \mu m$ \\
& $x \in [-525, 525]\,a_\text{HO} = [-1087, 1087]$ $\mu m$ (large amplitude) \\
Grid points & $N_x = 61{,}440$ \\
& $N_x = 92{,}160$ (large amplitude) \\
Grid spacing & $\Delta x \approx 0.0114\,a_\text{HO} \approx 0.024 \mu m$ \\
Timestep & $\Delta t = 5 \times 10^{-7}\,\omega_0^{-1} \approx 2.84 \times 10^{-9}$ s \\
Total evolution & 16 cycles shaking + 16 cycles hold = 32 cycles total \\
& (0.80--1.28 s depending on drive frequency) \\
Snapshots saved & 2000 \\
\midrule
Drive frequencies & $f_{\text{drive}} = 25, 30, 35, 40$ Hz \\
Drive amplitudes & 20 values, 0--180 $\mu m$ \\
& 6 values, 200--700 $\mu m$ (25 Hz only) \\
Atom numbers & $N_0 = 250{,}000$ (all frequencies) \\
& $N_0 = 1, 40{,}000, 100{,}000$ (40 Hz, 0--180 $\mu m$) \\
\bottomrule
\end{tabular}
\captionof{table}{Computational parameters for the split-operator simulations.}
\label{tab:parameters}
\end{center}

\vspace{0.3cm}

In our experiment the condensate \textcolor{black}{typically }contains $N_0 = 250{,}000$ atoms, but for comparison we also compute remaining-fraction curves for several other atom numbers. \textcolor{black}{Related data from experiments using condensates of $N_0=100,000$ and $40,000$ atoms are presented in Fig. \ref{fig:VariedAtomNumberODs}, as an experimental check that we understand the role of atom number and interatomic interactions in shaping the dynamics}. Varying $N_0$ shifts the chemical potential and provides additional physical intuition, much like how ionization yields in atomic systems depend on whether the valence-electron ionization potential becomes deeper or shallower.

The collective ground state $\psi_0(x)$ is obtained by imaginary-time evolution
($t \rightarrow -\mathrm{i}\tau$) using the same fourth-order Yoshida scheme with
$\Delta\tau = 5 \times 10^{-7}\,\omega_0^{-1}$, with renormalization after each step.
The evolution is continued until the chemical potential converges to the desired accuracy.
For the noninteracting case $N_0 = 1$—a deeply bound level in the Gaussian trap—we require
the converged energy to agree with the harmonic-oscillator ground-state energy to well
within $1\%$. For interacting condensates, the evolution is continued until the chemical
potential matches the corresponding Thomas--Fermi ground-state value (discussed in a later
section) to significantly better than $1\%$.

The chemical potential,
\begin{equation}
    \mu = \langle T \rangle + \langle V_{\text{trap}} \rangle + 2\langle U \rangle,
\end{equation}
is monitored throughout the evolution, and convergence is identified by its approach to a
steady value. For the parameters used, the converged ground states yield
$\mu \approx -303.2508\,\hbar\omega_0$ for $N_0 = 250{,}000$,
$\mu \approx -500.1081\,\hbar\omega_0$ for $N_0 = 40{,}000$, and
$\mu \approx -428.7398\,\hbar\omega_0$ for $N_0 = 100{,}000$.
The corresponding GPE densities $\rho_0(x) \equiv N_0 |\psi_0(x)|^2$ are shown in
Fig.~\ref{fig:GPE_HO_TF_ground_state}.

For the $N_0 = 1$ case, the converged value $\mu = -587.022\,\hbar\omega_0$
agrees closely with the truncated harmonic-oscillator ground-state energy
$\hbar\omega_0/2 - V_0 \approx -587.024\,\hbar\omega_0$.
Figure~\ref{fig:GPE_HO_TF_ground_state} also shows the corresponding density profile and its
comparison with the harmonic-oscillator ground state, with only minor numerical deviations
near $x \approx 0$ due to finite grid precision.

Stabilization is quantified through the remaining fraction, obtained by tracking the time-dependent particle number $N(t) = N_0 \int |\psi(x,t)|^2 dx$, with the wavefunction normalized such that $\int |\psi(x,0)|^2 dx = 1$. The remaining fraction is
\begin{equation}
\mathrm{Remaining}(t) = \frac{N(t)}{N_0}=\int|\psi(x,t)|^2dx,
\end{equation}
and is evaluated after 16 cycles of propagation in the shaking field, followed by 16 additional cycles without shaking (32 cycles total) to allow the majority of ejected atoms to exit the computational domain. Energy components (per atom) are computed as
\begin{align}
\langle T \rangle &= \int \psi^*(x)\left[-\frac{\hbar^2}{2m}\frac{\partial^2}{\partial x^2}\right]\psi(x)\,dx, \\
\langle V_\text{trap} \rangle &= \int V_{\text{trap}}(x)|\psi(x)|^2\,dx, \\
\langle U \rangle &= \frac{g(N_0-1)}{2}\int |\psi(x)|^4\,dx \overset{N_0\gg 1}{\approx} \frac{gN_0}{2}\int |\psi(x)|^4 dx,
\end{align}
and the kinetic energy is evaluated by applying the derivative operator in momentum space (via the inverse Fourier transform),
\begin{equation}
-\frac{\hbar^2}{2m}\frac{\partial^2\psi}{\partial x^2}
= \mathcal{F}^{-1}\!\left[\frac{\hbar^2 k^2}{2m}\tilde{\psi}(k)\right].
\end{equation}

\begin{figure}[t]
    \centering
    \includegraphics[width=\linewidth]{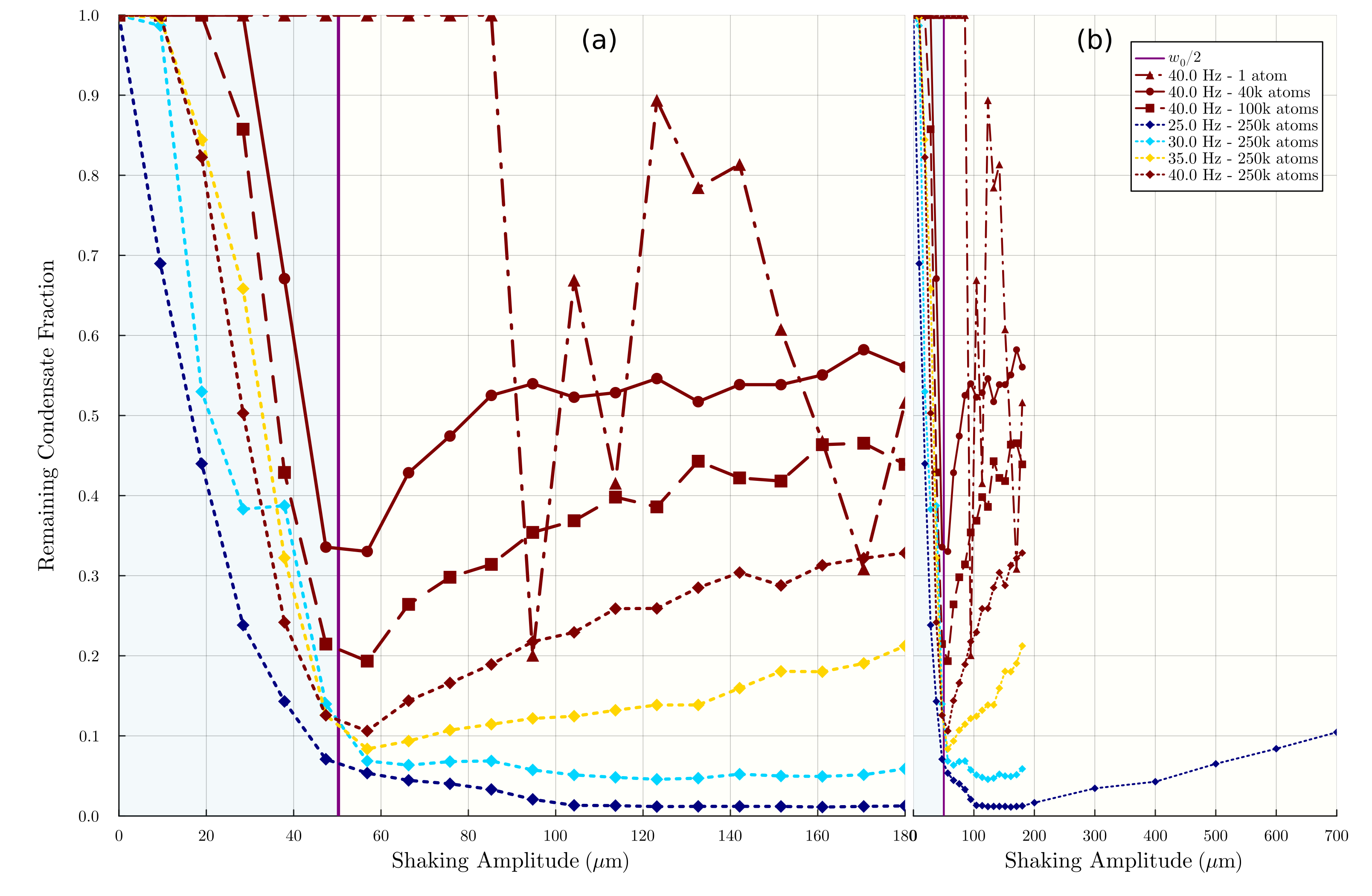}
    \caption{(a) Remaining fraction after 32 cycles (16 cycles of shaking + 16 cycles of rest) for $N_0 = 1$, $40{,}000$, $100{,}000$, and
    $250{,}000$ atoms for all simulated driving frequencies ($f_{\text{drive}} = 25$, $30$,
    $35$, and $40$ Hz). All shaking amplitudes up to $180~\mu$m are shown for every frequency.
    (b) Same data as in panel (a), but with the $25$ Hz curve extended to $700~\mu$m to provide
    a zoomed‑out view of the continued rise in the remaining fraction at large amplitudes.}

    \label{fig:remaining_fraction}
\end{figure}

Figure~\ref{fig:remaining_fraction} compiles the remaining-fraction data for atom numbers ranging from the single-particle limit ($N_0 = 1$) to the strongly interacting Thomas--Fermi regime ($N_0 = 40{,}000$, $100{,}000$, and $250{,}000$). The main panel shows simulations for $N_0 = 40{,}000$, $100{,}000$, and $250{,}000$ atoms at a driving frequency of $f_{\text{drive}} = 40~\text{Hz}$ with shaking amplitudes up to $180~\mu\text{m}$. To compare directly with our experimental conditions, we also include $N_0 = 250{,}000$ atom simulations at $f_{\text{drive}} = 25$, $30$, and $35~\text{Hz}$. For the slowest drive of $25~\text{Hz}$, the amplitude sweep is extended to $700~\mu\text{m}$ (shown in the inset), revealing that the apparent saturation at $180~\mu\text{m}$ is not a true plateau: the remaining fraction continues to increase monotonically with shaking amplitude.

Across all datasets, a pronounced minimum in the remaining fraction appears when the shaking amplitude approaches $a_0 \approx w/2$ for driving frequencies faster than the unshaken trap frequency, producing a clear “death valley’’ in the response. This minimum corresponds to the onset of strong depletion that occurs once the condensate can no longer adiabatically follow the rapidly moving trap. At still larger amplitudes, the remaining fraction begins to increase again, indicating the emergence of stabilization. For slower shaking, both the initial depletion and the subsequent stabilization are delayed: the remaining fraction decreases more gradually and recovers only at substantially larger amplitudes. 

Stabilization is observed for all atom numbers, including the single-atom case. The three interacting condensates ($N_0 = 40$k, $100$k, and $250$k) exhibit qualitatively similar behavior, with the onset of stabilization occurring at comparable shaking amplitudes. Quantitatively, the $40$k and $100$k datasets show reduced loss relative to the $250$k case. This is consistent with the fact that smaller condensates do not occupy as much of the trap volume: they sit lower in the potential and therefore spill fewer atoms when the trap is shaken. As a result, depletion is less severe and the recovery of the remaining fraction is somewhat more pronounced.

The single-atom calculation ($N_0 = 1$) also displays clear stabilization, demonstrating that the effect does not rely on mean-field interactions. However, the behavior of the remaining fraction differs substantially from the many-body case. In the absence of interactions, the remaining fraction shows pronounced oscillations as the peak shaking amplitude is increased, reflecting coherent single-particle sloshing in the shaken potential. By contrast, the interacting condensates—deep in the Thomas--Fermi regime—exhibit only very small residual oscillations. Their larger spatial extent and collective hydrodynamic response smooth the dependence on shaking amplitude, producing a much flatter remaining-fraction curve than in the single-particle limit.

A detailed physical interpretation of this sequence—where the breakdown of adiabatic following produces strong depletion at intermediate amplitudes, while the same supersonic trap motion ultimately enables stabilization at larger amplitudes—is developed in later sections of this appendix. These analyses focus specifically on the 40~Hz datasets for $N_0 = 250{,}000$ and $40{,}000$ atoms, which together span the full range of behaviors observed in Fig.~\ref{fig:remaining_fraction}. In Sec.~\textcolor{black}{B}\ref{sec:FFT_appendix_hydro}, a hydrodynamic analysis shows that the condensate’s ability to track the shaken trap is limited by the local speed of sound, and that the loss of adiabatic following coincides with the onset of depletion near $a_0 \approx w/2$. At larger amplitudes, however, the trap’s rapid motion causes it to spend most of its time near its turning points, reducing the effective forcing and allowing stabilization to emerge. In Sec.~\textcolor{black}{B}\ref{sec:FFT_appendix_bohm}, Bohmian trajectories provide a complementary picture: portions of the wavefunction are repeatedly ejected and recaptured when the trap moves too quickly to follow, while stabilized trajectories at large amplitudes exhibit negligible drift momentum. Sec.~\textcolor{black}{B}\ref{sec:slow_driving} then examines why stabilization persists even when the trap is shaken more slowly than the trap frequency, showing that large-amplitude motion alone is sufficient to recover stabilization provided that the envelope ramps up rapidly enough—set by the peak shaking amplitude and the finite number of drive cycles—to recapture populations that would otherwise be lost. Finally, Sec.~\textcolor{black}{B}\ref{sec:one_atom} analyzes the $N_0 = 1$ limit, where the dynamics reduce to pure single-particle stabilization, highlighting the fundamental differences between this non-interacting behavior and the collective, many-body stabilization observed in the experiment. We conclude this appendix in Sec.~\textcolor{black}{B}\ref{sec:strong-field_emulation} by connecting the hydrodynamic limits observed in our experiment to analogous limits in atomic and nuclear physics, and by speculating on how a shaken condensate may be used to emulate strong-field, non-perturbative responses in these heavier systems.

\textcolor{black}{
The motion of our condensate in the shaken trap provides a faithful analogue of electronic dynamics viewed in the non‑inertial Kramers–Henneberger frame, which follows the classical quiver trajectory of a free electron driven by a laser field~\cite{Argüello-Luengo_Rivera-Dean_Stammer_Maxwell_Weld_Ciappina_Lewenstein_2024}. 
Shaking the trap with amplitude $a_0$ corresponds to periodically driving the system with an external force
\begin{equation}
    F_{\mathrm{ext}} = m \omega_{\mathrm{drive}}^{2} a_0 .
\end{equation}
The maximum restoring force of the Gaussian trapping potential occurs at $x = w/2$, where
\begin{equation}
    F_{\mathrm{trap}} 
    = \frac{2 V_0 e^{-1/2}}{w}
    = \frac{1}{2} m \omega_0^{2} w e^{-1/2}.
\end{equation}
When $F_{\mathrm{ext}} < F_{\mathrm{trap}}$, the combined potential supports a barrier that confines particles, analogous to the tunneling‑ionization regime in atoms.  
In contrast, $F_{\mathrm{ext}} \ge F_{\mathrm{trap}}$ suppresses the barrier entirely, allowing population to flow directly into the continuum without tunneling—an exact analogue of over‑the‑barrier ionization.
}

\textcolor{black}{
The boundary between these two regimes is therefore set by the critical shaking amplitude required for barrier‑suppression ionization, $a_{\mathrm{BSI}}$, which satisfies
\begin{equation}\label{eqn:barrier_removal}
    \beta = \beta_{\mathrm{BSI}} \equiv \frac{a_{\mathrm{BSI}}}{w}
    = \frac{e^{-1/2}}{2}
      \left( \frac{\omega_0}{\omega_{\mathrm{drive}}} \right)^{2}
    \approx 0.3
      \left( \frac{\omega_0}{\omega_{\mathrm{drive}}} \right)^{2}.
\end{equation}
Across all shaking parameters used in this work, we find $\beta_{\mathrm{BSI}} \lesssim 0.38$.  
For $\beta < \beta_{\mathrm{BSI}}$, the combined potential retains a confining barrier, whereas for $\beta \ge \beta_{\mathrm{BSI}}$ the barrier is fully suppressed and over‑the‑barrier escape occurs.  
Bifurcations of the time‑averaged potential—and the onset of stabilization—appear only once $\beta \ge \beta_{\mathrm{c}} \equiv 8/9 - \varepsilon \approx 8/9$, which is strictly larger than $\beta_{\mathrm{BSI}}$.  
This progression, in which barrier suppression precedes the emergence of Kramers–Henneberger‑type stabilization, directly mirrors the sequence identified in low‑frequency strong‑field stabilization studies~\cite{volkova_ionization_2011}.
}

\section{Hydrodynamic Interpretation}\label{sec:FFT_appendix_hydro}
\textcolor{black}{A hydrodynamic formulation provides the natural dynamical scales needed to understand the response of the condensate under strong driving. }
The Gross--Pitaevskii equation (GPE) can be recast in the Madelung hydrodynamic form~\cite{Madelung1927} by writing
\begin{equation}
    \psi(x,t) = \sqrt{\rho(x,t)/N_0}\,e^{(\mathrm{i}/\hbar)S(x,t)},
\end{equation}
where $\rho(x,t)=N_0|\psi(x,t)|^2$ is the 1D density (with $\int |\psi|^2\,dx = 1$) and 
$v(x,t)=\frac{1}{m}\partial_x S$ is the velocity field.  
This decomposition yields the continuity equation
\begin{equation}\label{eqn:continuity}
    \frac{\partial \rho}{\partial t}
    + \frac{\partial j}{\partial x}
    = -\frac{\Gamma}{\hbar}\rho,
\end{equation}
and the Hamilton--Jacobi equation
\begin{equation}
    -\frac{\partial S}{\partial t}
    = \frac{(\partial_x S)^2}{2m}
    + V_{\text{trap}}
    + V_{\text{quantum}}
    + V_{\text{classical}},
\end{equation}
where $j=\rho v = N_0\frac{\hbar}{m}\mathrm{Im}[\psi^*\partial_x\psi]$ is the current.  
The quantum-pressure and mean‑field interaction potentials are
\begin{equation}
    V_{\text{quantum}} = -\frac{\hbar^2}{2m}\frac{\partial_x^2\sqrt{\rho}}{\sqrt{\rho}},
    \qquad
    V_{\text{classical}} = g\rho.
\end{equation}

Taking the spatial derivative of the Hamilton--Jacobi equation and using $v=\frac{1}{m}\partial_x S$ yields the Euler equation
\begin{equation}\label{eqn:Euler}
    \frac{\partial v}{\partial t}
    + v\,\frac{\partial v}{\partial x}
    = -\frac{1}{m}\frac{\partial}{\partial x}
    \!\left[V_{\text{trap}} + V_{\text{quantum}} + V_{\text{classical}}\right].
\end{equation}
Here the classical ``pressure'' term represents an internal bulk pressure associated with repulsive interactions in regions of high density, while the quantum‑pressure term acts as a stabilizing surface‑tension‑like contribution associated with density gradients.

For the interacting GPE datasets, the ground-state density is well described by the
Thomas--Fermi (TF) approximation. Neglecting quantum pressure in the stationary
Hamilton--Jacobi equation gives
\begin{equation}
    \mu \approx V_{\text{trap}}(x) + g\rho_0(x),
\end{equation}
from which the TF density profile follows:
\begin{equation}
    \rho_0(x) \approx \frac{\mu - V_{\text{trap}}(x)}{g}\,
    \theta\!\left(\mu - V_{\text{trap}}(x)\right),
\end{equation}
with the TF radius $a_{\text{TF}}$ defined implicitly by $\mu = V_{\text{trap}}(\pm a_{\text{TF}})$.

Figure~\ref{fig:GPE_HO_TF_ground_state} compares the Thomas--Fermi (TF) prediction with the full
GPE ground-state density for $N_0 = 40{,}000$, $100{,}000$, and $250{,}000$ atoms. Although the
TF approximation neglects quantum‑pressure (surface‑tension) effects, the resulting
discrepancies are small relative to the overall cloud size, and the corresponding errors in the
chemical potential (values shown in the figure) remain well below $1\%$. The GPE solution
smooths the TF edges through quantum‑pressure effects, but the close agreement across these
three atom numbers confirms that these interacting initial states lie firmly within the
Thomas--Fermi regime.

Interestingly, both the $N_0 = 1$ harmonic‑oscillator–like density and the larger‑$N_0$
Thomas--Fermi densities exhibit an approximately Gaussian central profile. However, the
$N_0 = 1$ state is Gaussian globally, whereas the TF profiles for large $N_0$ fall sharply to
zero at the condensate edges.

Figure~\ref{fig:unshaken} further illustrates this by showing the TF fillings for each atom
number—defined as the portion of the unshaken trap occupied in energy up to the corresponding
chemical potential—providing a direct visualization of how each condensate populates the static
Gaussian potential. This behavior contrasts sharply with earlier theoretical studies of weakly
interacting condensates~\cite{Argüello-Luengo_Rivera-Dean_Stammer_Maxwell_Weld_Ciappina_Lewenstein_2024},
which examined the opposite limit: shaking in a quantum‑pressure‑dominated regime where
$s$‑wave interactions play only a minor role.

The hydrodynamic formulation provides the natural dynamical scales needed to understand the response of the condensate under strong shaking. In particular, the speed of sound sets the rate at which density perturbations propagate through the fluid \cite{pitaevskii2016bose}. In general,
\begin{equation}
    c^2 = \frac{1}{m}\left(\frac{\partial P}{\partial \rho}\right),
\end{equation}
where $P$ is the pressure. For a dilute Bose gas with equation of state $P=\tfrac{1}{2}g\rho^2$, this reduces to
\begin{equation}
    c = \sqrt{\frac{g\rho_0}{m}}.
\end{equation}

Linearizing the continuity and Euler equations around a static TF background 
$\rho(x,t)=\rho_0+\delta\rho$ and $v=\delta v$, and neglecting quantum pressure (valid in the TF regime), gives
\begin{equation}
    \frac{\partial\,\delta\rho}{\partial t}
    + \rho_0\,\frac{\partial}{\partial x}\delta v = 0,
\end{equation}
\begin{equation}
    \frac{\partial\,\delta v}{\partial t}
    = -\frac{g}{m}\,\frac{\partial}{\partial x}\delta\rho,
\end{equation}
where our analysis focuses on flow far from the absorbing region ($|x|$ small enough that $\Gamma(x)=0$). Eliminating $\delta v$ yields a wave equation for density fluctuations,
\begin{equation}
    \frac{\partial^2\,\delta\rho}{\partial t^2}
    = \frac{g\rho_0}{m}\,\frac{\partial^2\,\delta\rho}{\partial x^2},
\end{equation}
confirming that the coefficient of the spatial derivative is $c^2$.

At the trap center, where $\rho_0(0) = (\mu + V_0)/g$, this becomes
\begin{equation}
    c = \sqrt{\frac{\mu + V_0}{m}}.
\end{equation}
For the ground states used here, this gives  
$c \approx 16.9\,a_\text{HO}\omega_0$ (6.2 mm/s) for $N_0=250{,}000$,  
$c \approx 9.4\,a_\text{HO}\omega_0$ (3.4 mm/s) for $N_0=40{,}000$, and  
$c \approx 12.7\,a_\text{HO}\omega_0$ (4.6 mm/s) for $N_0=100{,}000$.

\begin{figure}[H]
    \centering
    \includegraphics[width=0.95\linewidth]{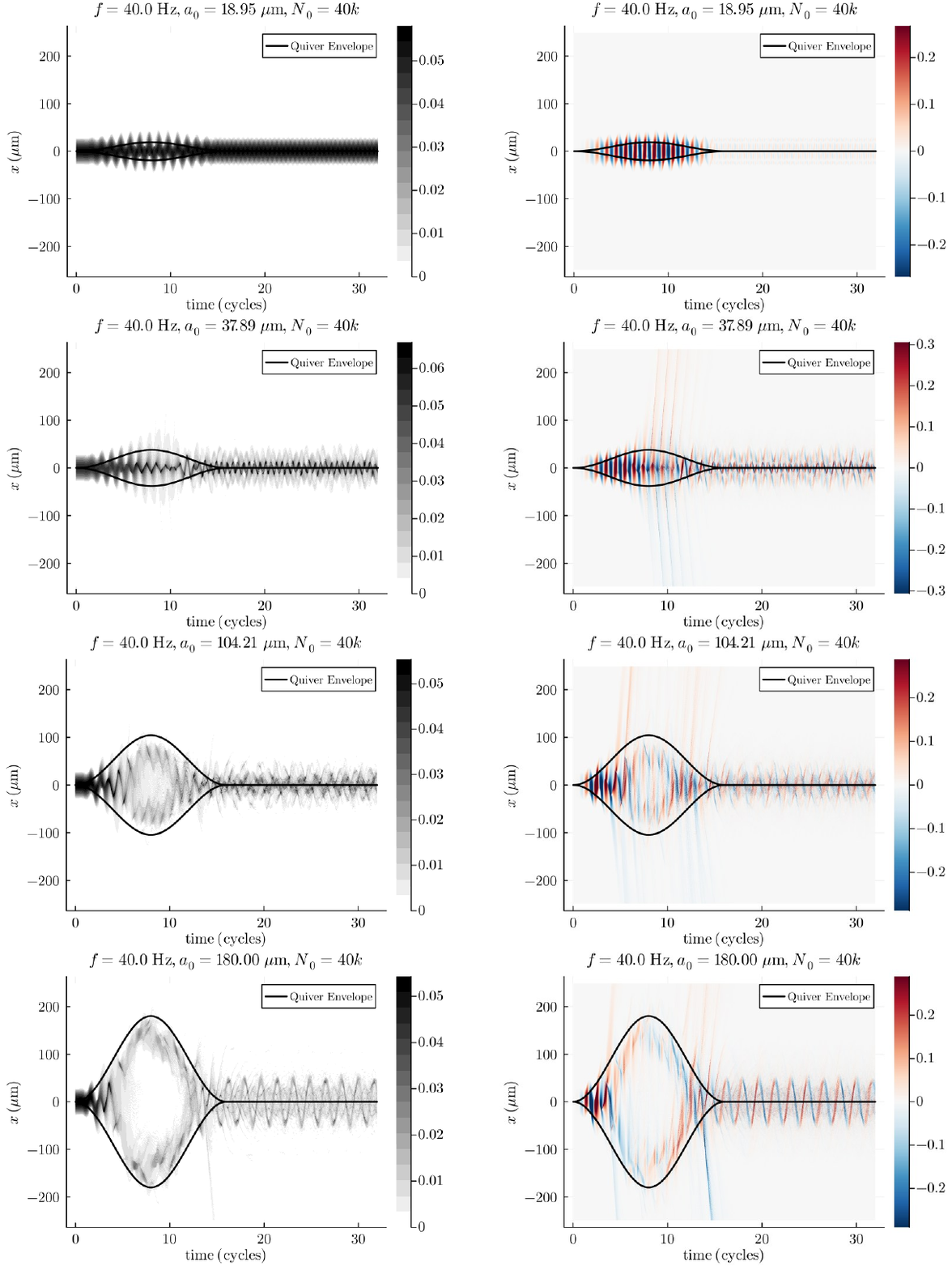}
    \caption{Condensate density (left) and current density (right) for 
    $N_0=40{,}000$ atoms at 40~Hz. Each row corresponds to a different 
    shaking amplitude: 18.95~µm, 37.89~µm, 104.21~µm, and 180~µm. 
    All densities and current densities are shown in harmonic‑oscillator units. 
    The positive and negative shaking envelopes are overlaid as black curves in each 
    subplot, illustrating how the amplitudes of the density oscillations and the 
    stabilized condensate remain approximately bounded by the envelope.}
    \label{fig:40k_densities}
\end{figure}

\begin{figure}[H]
    \centering
    \includegraphics[width=0.95\linewidth]{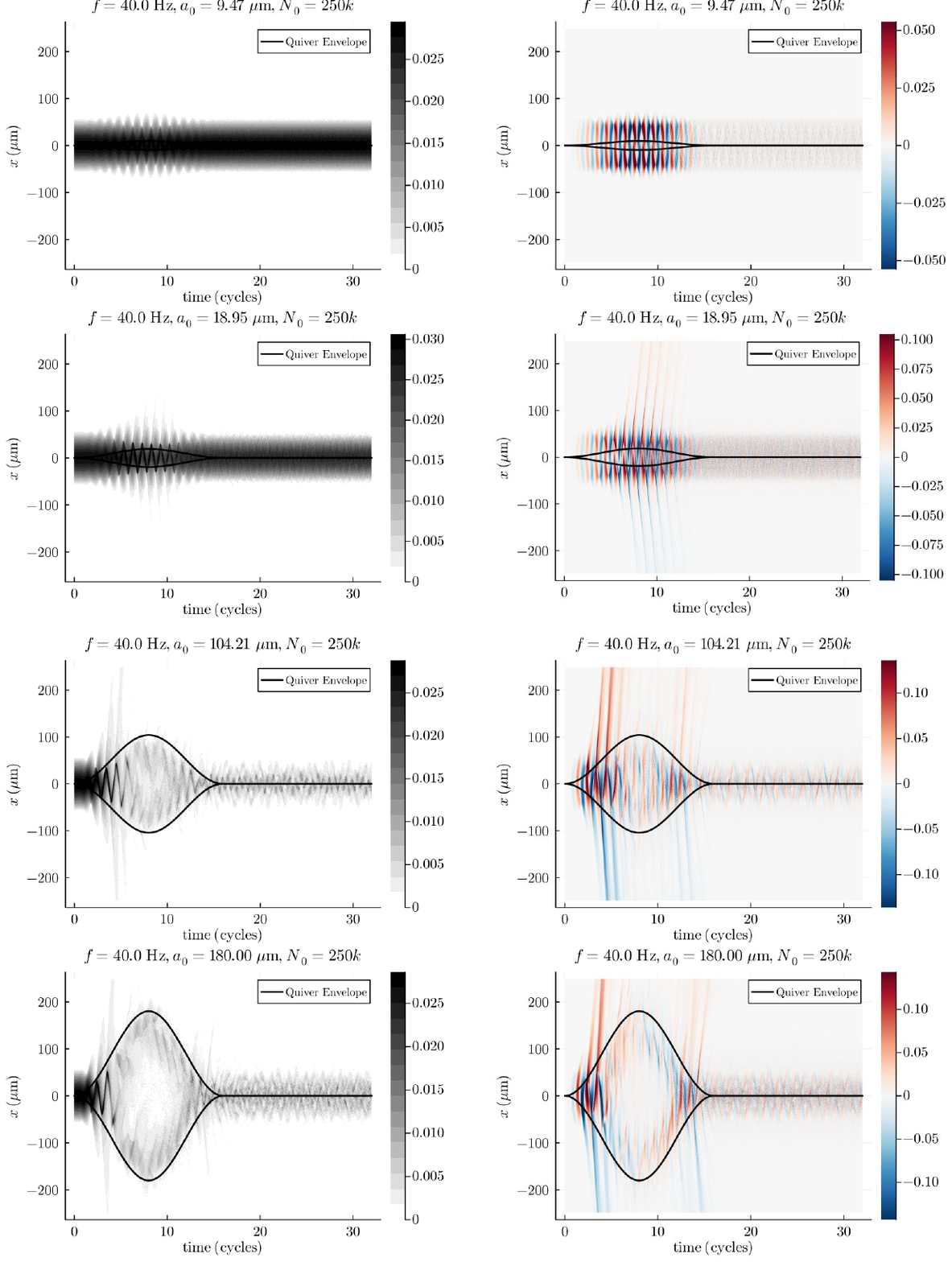}
    \caption{Condensate density (left) and current density (right) for 
    $N_0=250{,}000$ atoms at 40~Hz. Shaking amplitudes are 9.47~µm, 
    18.95~µm, 104.21~µm, and 180~µm. 
    All densities and current densities are shown in harmonic‑oscillator units. 
    As in Fig.~\ref{fig:40k_densities}, the positive and negative shaking envelopes are 
    plotted as black curves within each subplot, showing that the density oscillations and 
    stabilized profiles occupy a region approximately bounded by the envelope.}
    \label{fig:250k_densities}
\end{figure}

Since the trap center moves as $x_{\text{trap}}(t)=a(t)$, the trap velocity is
\begin{equation}
    v_{\text{trap}}(t) = \frac{d}{dt}a(t),
\end{equation}
with maximum value (for $a(t)=a_0(t)\sin(\omega_{\text{drive}} t)$)
\begin{equation}
    v_{\text{max}} \approx a_0(t)\,\omega_{\text{drive}},
\end{equation}
where this approximation assumes that the envelope varies slowly compared to the drive frequency ($a_0\omega_{\text{drive}}\gg \partial_t a_0(t)$). The motion becomes supersonic when
\begin{equation}
    a_0 \gtrsim \frac{c}{\omega_{\text{drive}}},
\end{equation}
which, depending on atom number and drive frequency, ranges from 14--39\,$\mu m$.  
When the trap moves faster than the condensate can hydrodynamically redistribute, steep density gradients develop and the condensate cannot follow the instantaneous trap minimum.  
Substantial distortion is expected when the shaking amplitude also exceeds the Thomas–Fermi radius
\begin{equation}
    a_{\text{TF}} = \frac{w}{\sqrt{2}}
    \sqrt{\log\!\left(\frac{V_0}{|\mu|}\right)}.
\end{equation}

Supersonic motion ensures that information cannot propagate across the condensate fast enough to maintain equilibrium, and while this guarantees hydrodynamic stress and steep gradients, it does not by itself imply global deformation. Interactions give the condensate a form of stiffness, and the sound speed characterizes how quickly density waves propagate. In the limit $g \to 0$, the sound speed $c \to 0$, reflecting the absence of any restoring force for density perturbations. In reality $c$ does decrease with $g$. However, we take care to assume that $gN_0$ is not taken exactly to zero as we assumed the hydrodynamic Thomas-Fermi limit when obtaining $c$.

With these dynamical scales established, we now examine the shaken‑trap dynamics for 
$N_0=40{,}000$ and $250{,}000$ atoms at 40~Hz. 
Figures~\ref{fig:40k_densities} and \ref{fig:250k_densities} show the condensate density 
(left panels) and current density (right panels), with densities and currents plotted in 
harmonic‑oscillator units. In each subplot, the positive and negative shaking envelopes are 
overlaid as black curves, illustrating how the density oscillations and the stabilized 
condensate occupy a region approximately bounded by the envelope. 
Figure~\ref{fig:40k_densities} corresponds to the $40{,}000$‑atom condensate, while 
Fig.~\ref{fig:250k_densities} shows the $250{,}000$‑atom condensate. In each figure, the 
four rows represent a progression from bound‑state dynamics (top row), to strong‑field loss 
(second row), to weak stabilization (third row), and finally to strong stabilization 
(bottom row).

In the top row, the shaking amplitude is small enough that the trap velocity remains
subsonic or only weakly supersonic. For the 250k-atom condensate, $9.47~\mu\mathrm{m}$ lies
well below the sound speed ($c_{250k} \approx 6.2~\mathrm{mm/s}$), and the condensate follows
the trap minimum with only mild distortion. For the 40k-atom condensate, the corresponding
threshold is $a_0 \approx 14~\mu\mathrm{m}$, so $18.95~\mu\mathrm{m}$ is already slightly
supersonic ($c_{40k} \approx 3.4~\mathrm{mm/s}$), but the motion still resembles a bound
state with limited deformation. In both cases, the condensate can hydrodynamically
redistribute quickly enough to remain confined, and the density oscillates collectively with
the trap.

The second row shows the onset of strong-field loss. Here the trap velocity exceeds the
local sound speed, and the condensate can no longer redistribute rapidly enough to remain
bound. For the 250k-atom condensate this occurs at $18.95~\mu\mathrm{m}$, while for the
40k-atom condensate both $18.95~\mu\mathrm{m}$ and especially $37.89~\mu\mathrm{m}$ lie in
the supersonic regime. This produces pronounced compression of the density profile: the
central region becomes significantly darker in the density plots, and the maximum density
reported in the colormap increases noticeably compared to the top row. These higher-density
regions reflect the fluid being driven into a narrower spatial region before being expelled.
In this regime, bursts of density escape the trap every half-cycle, closely resembling
strong-field ionization. These amplitudes remain smaller than $w/2$ and therefore do not
yet produce stabilization.

The third row corresponds to the onset of stabilization, but in a weak and incomplete form.
At $104.21~\mu\mathrm{m}$, the instantaneous trap velocity is already far above the
sound-speed scale, so the condensate cannot follow the carrier oscillations. As a result, the
strong-field loss seen in the second row is greatly reduced: the rapid shaking no longer
compresses the density into the steepest part of the Gaussian each half-cycle. However, the
time-averaged potential is only weakly developed at this amplitude. The central barrier in
the phase-averaged potential remains shallow, and the bifurcated minima are not yet deep or
wide enough to fully confine the density. Consequently, the condensate can still flow over
the central barrier with little resistance, especially during the envelope ramp-up and
ramp-down, which induce slow, large-scale motion that the fluid can respond to. The density
therefore reflects a mixture of carrier-scale and envelope-scale dynamics: the carrier no
longer produces significant forces, while the envelope dominates the residual motion. This
produces a weak form of stabilization—loss is suppressed, but the condensate is not yet
locked into the fully developed time-averaged wells.

The bottom row shows strong stabilization at $180~\mu\mathrm{m}$. Here the shaking amplitude
is far above the sound-speed scale, and the time-averaged potential is fully developed: the
effective double-well becomes wide, and the central barrier grows tall relative to the well
depth. The condensate density thins dramatically in the barrier region and accumulates near
the bifurcated minima, where the trap velocity is nearly zero. Although the condensate is
stabilized on average, some loss still occurs during the transonic portions of the envelope
ramp-up and ramp-down, when the instantaneous amplitude passes through the same weakly
supersonic range that produces strong-field loss in the second row. The stabilization
behavior is qualitatively similar for both atom numbers, but the 250k-atom condensate fills
the time-averaged trap more strongly due to its higher density, which allows it to occupy
regions of higher energy; the trap acts like a vessel that is more highly filled.

The response of the condensate to the shaken trap is naturally organized into three regimes.
(1)~In the bound regime, which includes both subsonic motion and the weakly supersonic
velocities reached for the 40k condensate, the fluid remains confined and the current
oscillates collectively with the trap. (2)~As the drive amplitude increases, the trap motion
becomes fast enough to reposition the condensate across different regions of the Gaussian
beam, producing a strong-field loss regime. The most pronounced depletion occurs when the
shaking amplitude passes through $a_0 \approx w/2$, where the condensate is placed over the
steepest part of the Gaussian and the instantaneous force is largest. In this ``death-valley''
region the fluid is strongly compressed and most efficiently expelled, producing sharp
half-cycle bursts in the current. For amplitudes smaller than $w/2$, the condensate samples
a shallower region of the trap where the force is weaker and depletion is reduced. For
amplitudes significantly larger than $w/2$, the condensate is displaced onto the broad
wings of the trap where the force again decreases; atoms may still be ejected, but they
acquire only very small drift momenta because the local acceleration is weak. (3)~At still
larger amplitudes, where the trap velocity becomes far greater than the local sound speed,
the system enters a stabilization regime. Here the condensate can no longer follow the rapid
carrier oscillations and instead tracks the bifurcated minima of the phase-averaged
potential, and the current becomes locked to the slow envelope. As discussed in the following
section, a Bohmian analysis shows that trajectories in this regime are repeatedly ejected and
recollide with the trap every half cycle, yet these collisions do not significantly modify
their small drift velocities. Once the envelope ramps down and the trap becomes subsonic,
these weakly drifting trajectories are readily recaptured, leaving only residual loss from
the transonic portions of the envelope ramp where the instantaneous amplitude again passes
through $a_0 \approx w/2$.

\section{Microscopic Dynamics from Bohmian Trajectories}\label{sec:FFT_appendix_bohm}
The density and current–density plots in the previous section describe the collective, 
hydrodynamic response of the condensate to the shaken trap. To interpret the microscopic 
velocities of atoms that underlie those macroscopic features—particularly the origin of 
compression, excitation, ionization, and stabilization—we examine Bohmian trajectories. 
Originally introduced in the context of an alternative formulation of quantum mechanics 
\cite{bohm1952I,bohm1952II}, Bohmian trajectories have also been widely used in 
strong-field ionization studies as a practical way to visualize the flow of probability 
current \cite{sanz2013trajectory,douguet2018dynamics,moon2024strong}. Here we apply the same trajectory-based 
analysis to BEC loss from a shaken Gaussian trap.

Although Bohmian mechanics offers a literal trajectory‑based ontology for quantum dynamics, our 
use of trajectories does not depend on adopting that interpretation. For the purposes of this 
work, we treat the curves as representing the motion of individual atoms within the collective 
flow, in the same operational sense that Bohmian trajectories are used in strong‑field physics 
to visualize microscopic dynamics. This does not commit us to any particular ontology; it 
simply provides a convenient and intuitive way to track how different portions of the 
condensate move and separate in time, consistent with the density and current‑density 
evolution discussed in the previous section.

The particle velocity field is given by $v(x,t)=j(x,t)/\rho(x,t)$, and individual trajectories 
evolve according to the guidance equation
\begin{equation}
    \frac{dx}{dt}=v\big(x(t),t\big),
\end{equation}
which, together with the continuity equation (Eq.~\eqref{eqn:continuity}), ensures that an 
ensemble of trajectories initially distributed as $\rho(x,0)=N_0|\psi(x,0)|^2$ preserves the 
quantum distribution at all later times.

The GPE simulation stores $N_{\text{snap}}=2{,}000$ wavefunction snapshots spanning the full 
evolution (16 shaking cycles followed by 16 stationary cycles). For each snapshot we compute 
the velocity field
\begin{equation}
v(x,t)=\frac{\hbar}{m}\frac{\mathrm{Im}\!\left[\psi^*(x,t)\,\partial_x\psi(x,t)\right]}
{|\psi(x,t)|^2},
\end{equation}
with $\partial_x\psi$ evaluated spectrally via
\begin{equation}
\partial_x\psi=\mathcal{F}^{-1}[ik\,\tilde{\psi}(k)].
\end{equation}
To avoid numerical instabilities, velocities are computed only where 
$\rho(x,t)>10^{-4}\max[\rho(x,t)]$. The full set of velocity fields $\{v(x,t_i)\}$ is 
precomputed and stored within the plot region $|x|<x_{\text{limit}}$.

Trajectories are integrated using a fourth-order Runge--Kutta scheme with 
$N_{\text{sub}}=50$ substeps per snapshot interval. The velocity field at intermediate times 
is obtained by linear interpolation between snapshots, and spatial values are evaluated using 
cubic splines. Trajectories are terminated once they exceed $\pm 220\,\mu\mathrm{m}$ for
shaking amplitudes up to $180\,\mu\mathrm{m}$, and at $800\,\mu\mathrm{m}$
for larger-amplitude sweeps reaching as high as $700\,\mu\mathrm{m}$ discussed
in the following section. These cutoffs lie well outside both the
Thomas--Fermi radius and the effective trap region for the shaking parameters
used in our calculations.
 Initial positions are sampled 
deterministically from the cumulative distribution of the initial density, restricted to 
$\rho_0(x)>0.3\max[\rho_0(x)]$, ensuring that each of the $N_{\text{traj}}=1{,}000$ 
trajectories represents an equal fraction of the atom number.

Within this sampling region, we construct the cumulative distribution function
\begin{equation}
    P(x)=\frac{1}{\mathcal{Z}}\int_{x_{\min}}^{x}\rho_0(x')\,dx'
    \approx \frac{1}{\mathcal{Z}}\sum_{x_i\le x}\rho_0(x_i)\,\Delta x,
\end{equation}
where $x_{\min}$ and $x_{\max}$ denote the boundaries of the sampling region and 
$\mathcal{Z}\equiv\int_{x_{\min}}^{x_{\max}}\rho_0(x')\,dx'$ ensures $P(x_{\max})=1$.  
We then place the $N_{\text{traj}}=1{,}000$ initial positions at points $x_0^{(k)}$ satisfying 
$P(x_0^{(k)})=(k-1/2)/N_{\text{traj}}$ for $k=1,\dots,N_{\text{traj}}$, so that each trajectory 
represents an equal fraction of the atom number and the sampling is naturally concentrated near 
the trap center where the density is highest.

Although our ground state lies safely within the Thomas--Fermi limit, we retain the quantum--
pressure term in both the ground--state density and the time--dependent response. Bohmian 
trajectories in one dimension obey the fundamental property that they cannot cross: the 
velocity field is single--valued, and trajectories initialized at positions $x_a<x_b$ must 
satisfy $x_a(t)<x_b(t)$ for all later times \cite{bohm1952I,bohm1952II,holland1995quantum}. As a result, if two trajectories that begin at 
$x_a$ and $x_b$ remain bound to the trap throughout the evolution, then every trajectory 
initially located between them must also remain bound, since none can pass through the bounding 
trajectories. During ionization, trajectories furthest from the trap center therefore tend to 
be ejected first, with the loss front moving inward in a well-defined sequence \cite{sanz2013trajectory}. This is 
precisely the regime where surface tension (quantum pressure) becomes increasingly important: 
as $\rho_0(x)\to 0$ near the condensate edges, bulk pressure contributions diminish while 
gradient terms dominate \cite{dalfovo1999theory}.

Furthermore, bursts of ejected atoms (small ``droplets'') produced near the half--cycles of the 
driving field \cite{sanz2013trajectory} may contain any number of atoms, but their populations remain a small fraction of 
the total initial atom number $N_0$. In these small droplets, the density is low and the atoms 
are no longer confined by the trap, so bulk contributions are reduced and quantum pressure 
(surface tension) can play a comparatively larger role in shaping their dynamics.

\begin{figure}[H]
    \centering
    \includegraphics[width=\linewidth]{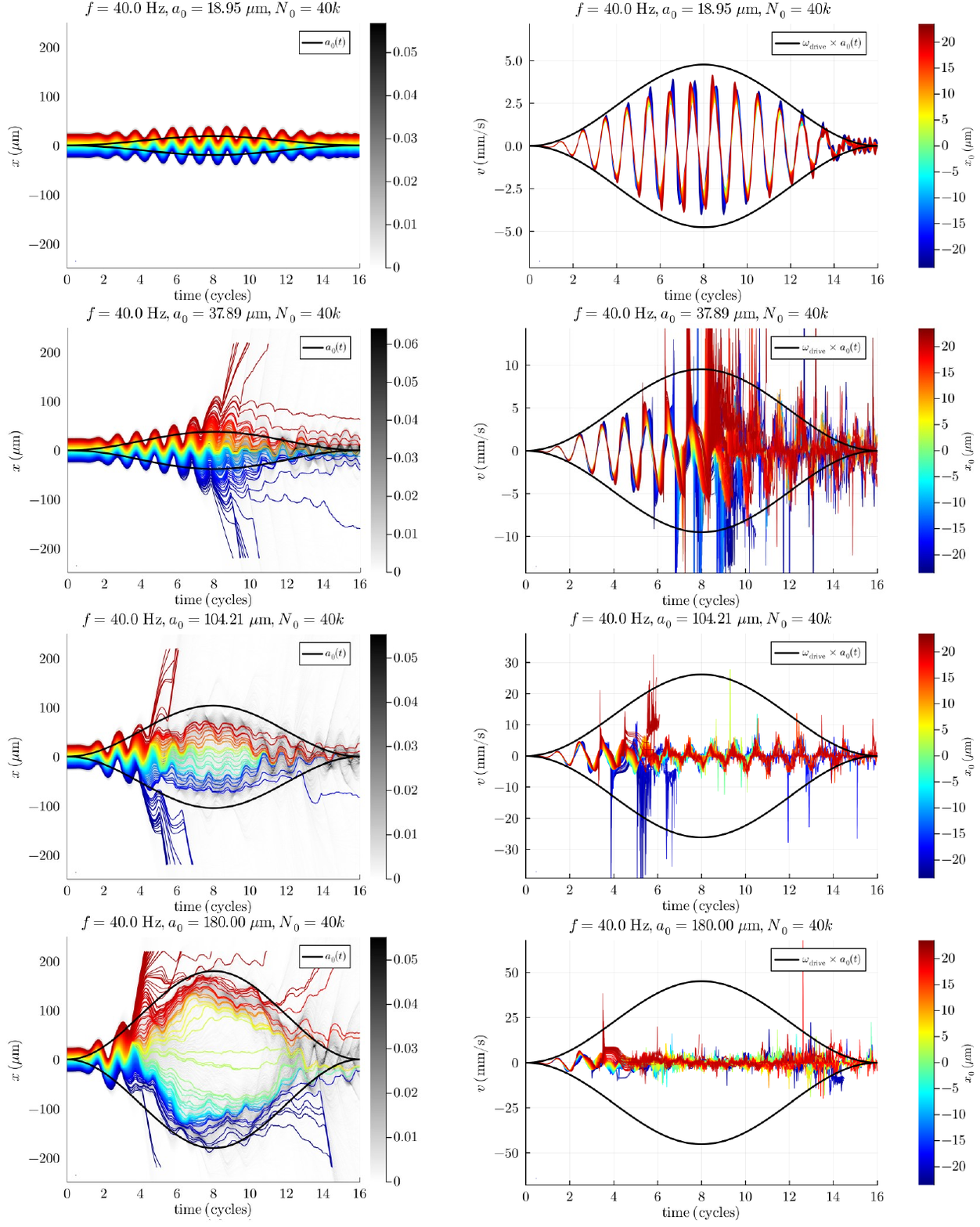}
    \caption{Bohmian trajectories for $N_0=40{,}000$ atoms at 40~Hz. 
    Left: density heatmaps with trajectories overlaid (black curves indicate the shaking envelope at $x=\pm a_0(t)$). 
    Right: corresponding Bohmian velocities (black curves indicate the velocity envelope at $v\approx \pm a_0(t)\omega_\text{drive}$). 
    Colors indicate initial coordinate $x_0$. 
    Rows correspond to the same shaking amplitudes as in Fig.~\ref{fig:40k_densities}.}
    \label{fig:40k_bohm}
\end{figure}

\begin{figure}[H]
    \centering
    \includegraphics[width=\linewidth]{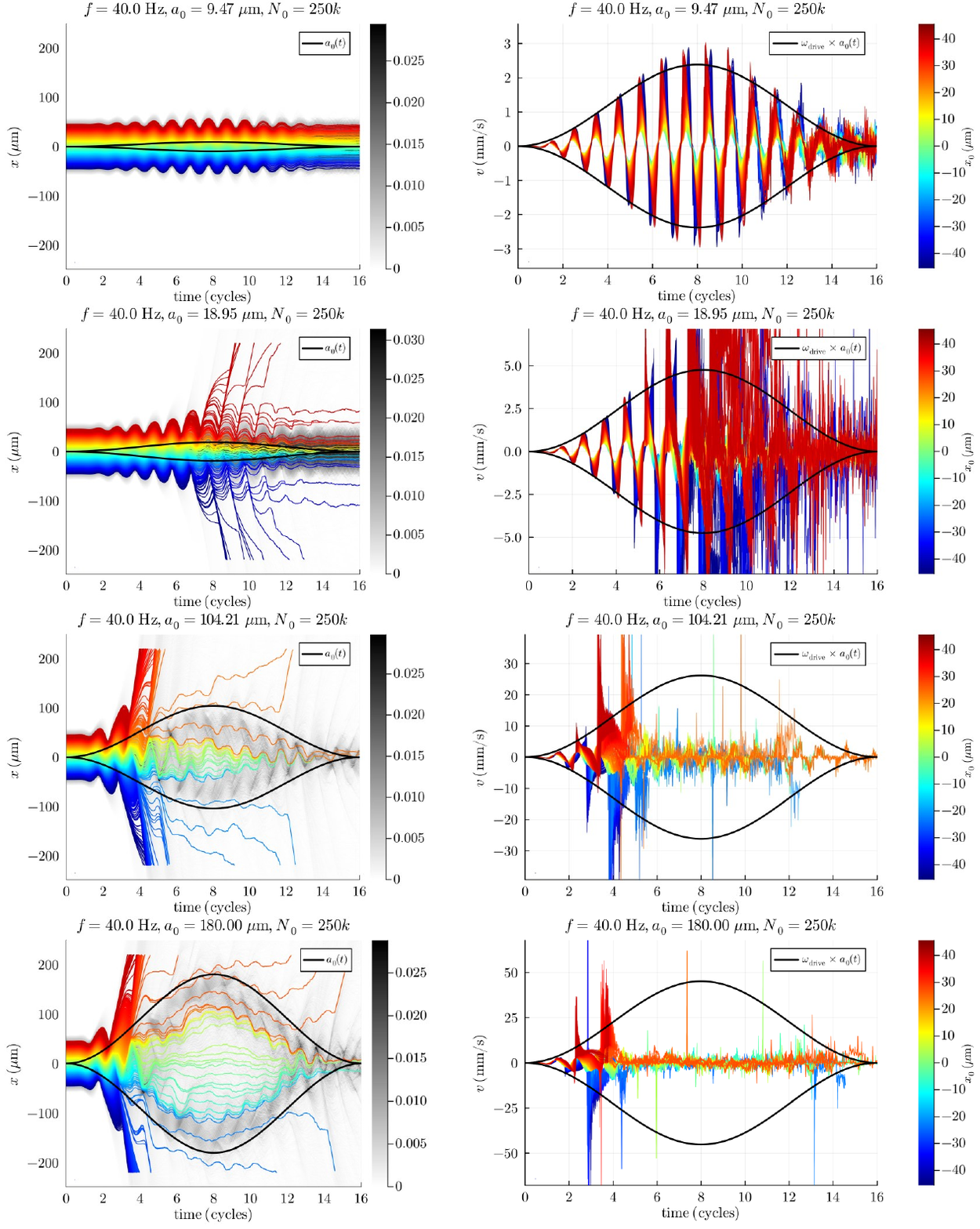}
    \caption{Bohmian trajectories for $N_0=250{,}000$ atoms at 40~Hz. 
    Left: density heatmaps with trajectories overlaid (black curves indicate the shaking envelope at $x=\pm a_0(t)$). 
    Right: corresponding Bohmian velocities (black curves indicate the velocity envelope at $v\approx \pm a_0(t)\omega_\text{drive}$).
    Colors indicate initial coordinate $x_0$. 
    Rows correspond to the same shaking amplitudes as in Fig.~\ref{fig:250k_densities}.}
    \label{fig:250k_bohm}
\end{figure}

Figures~\ref{fig:40k_bohm} and \ref{fig:250k_bohm} each contain two columns. The left column 
shows density heatmaps with Bohmian trajectories overlaid; the color bar corresponds to the 
density. The right column shows the corresponding Bohmian velocities, obtained by 
differentiating the trajectories in time. In both columns, the color of each trajectory is 
determined by its initial coordinate $x_0$: red trajectories originate on the positive 
(right–most) side of the condensate, green trajectories originate near $x=0$, and blue 
trajectories originate on the negative (left–most) side, with intermediate colors in between. 
The same color labeling is used consistently across density and velocity panels. As before, 
the shaking envelope is plotted as black curves in the left panels; in the right panels we 
plot $\omega_{\mathrm{drive}}$ times the envelope to provide an approximate velocity envelope.

The top row of each figure corresponds to the bound-state regime. Here the trajectories 
oscillate collectively with the trap. Red and blue trajectories begin in phase with the 
driving field but slowly dephase over time, most clearly visible in the velocity plots. 
Higher-frequency oscillations develop, analogous to higher harmonics in strong-field physics, 
and represent excitation of the condensate. For $N_0=250$k, the velocity oscillations nearly 
span the full velocity envelope $\omega_{\mathrm{drive}}a_0(t)$, while for $N_0=40$k they fall 
slightly short, consistent with the weakly supersonic regime where the condensate cannot fully 
keep up with the trap motion.

The second row corresponds to the strong-field loss regime. Trajectories initialized near the 
center of the trap (green, yellow, light blue) are sharply compressed as the envelope ramps 
up, while red and blue trajectories initialized near the edges spread outward and a fraction 
become ionized. Because Bohmian trajectories in one dimension cannot cross, this compression 
and spreading directly reflects the microscopic structure of the wavefunction. The 
corresponding velocities oscillate with the field at early times but become highly erratic 
after excitation due to large superpositions and ionization. Trajectories that propagate 
beyond $|x|>220\,\mu\mathrm{m}$ are terminated and no longer shown, reflecting atoms that have 
effectively escaped the trap. 

The third row shows the weak stabilization regime. Loss during the initial ramp-up resembles
the strong-field case, but loss is greatly reduced once $a_0$ exceeds $w/2$. Before
stabilization, the trajectory velocities oscillate with the trap velocity at the carrier
frequency. After stabilization sets in, however, the velocities collapse to values near zero,
with only small residual oscillations. In the velocity plots, the envelope
$a_0(t)\,\omega_{\mathrm{drive}}$ simply marks the extrema of the trap velocity; the
trajectories do not follow this envelope once stabilized. Between these brief turning-point
interactions, the atoms coast with almost no drift velocity. This makes them easy to
recapture as the trap slows during the ramp-down, though some loss still occurs in the
transonic region.

The bottom row shows the strong stabilization regime. Trajectories bunch near the minima of
the bifurcated time-averaged potential, reflecting the structure of the fully developed
effective double well. Ionization and re-scattering persist, but the trap moves so rapidly
that atoms interact with it primarily near the turning points where the trap velocity nearly
vanishes. These turning points dominate the microscopic dynamics and explain the stabilized
density profiles observed earlier.

Finally, it is worth noting a fundamental feature of Bohmian mechanics that helps interpret the trajectory plots presented here. Because the velocity field is single–valued and Bohmian trajectories in one dimension never cross \cite{bohm1952I,bohm1952II,holland1995quantum} each trajectory preserves its ordering relative to all others throughout the evolution. As a result, if one could detect the position of a single trajectory at some later time—whether bound, ejected, or stabilized—its entire past history would be uniquely determined by integrating the guidance equation backward in time \cite{holland1995quantum,wyatt2005quantum}. This backward integration would return the trajectory to a definite initial coordinate $x_0$, identifying the precise region of the ground–state density from which that atom originated. In this sense, every point in the final distribution carries a well–defined “label’’ inherited from its initial position in the condensate, and the Bohmian flow provides a one–to–one mapping between the initial density profile and the microscopic fate of each portion of the wavefunction. Although we use trajectories here only as a diagnostic tool rather than as an ontological commitment, this invertibility of the flow underlies their interpretive power: the microscopic structure of excitation, compression, ionization, and stabilization can be traced directly back to specific regions of the initial condensate \cite{sanz2013trajectory,douguet2018dynamics}.

\section{Stabilization in the Slow‑Driving Regime}\label{sec:slow_driving}

We now examine condensate dynamics in the slow‑driving regime, where the shaking frequency
($25\,\mathrm{Hz}$) lies well below the harmonic frequency of the trap. All simulations in this
section use $N_0 = 250{,}000$ atoms, and we explore shaking amplitudes up to
$a_0 = 700\,\mu\mathrm{m}$. As in the 40\,Hz analysis, we present density and current‑density
evolution together with Bohmian trajectories. Slow driving produces strong large‑scale motion
and substantial loss during the early cycles of the pulse, even at relatively small
amplitudes. However, at sufficiently large amplitudes the system eventually enters a
stabilized regime, with surviving trajectories becoming confined by the time‑averaged
potential after the initial loss window. The progression from early‑cycle loss to late‑cycle
stabilization unfolds gradually as the shaking amplitude increases.

\begin{figure}[H]
    \centering
    \includegraphics[width=\linewidth]{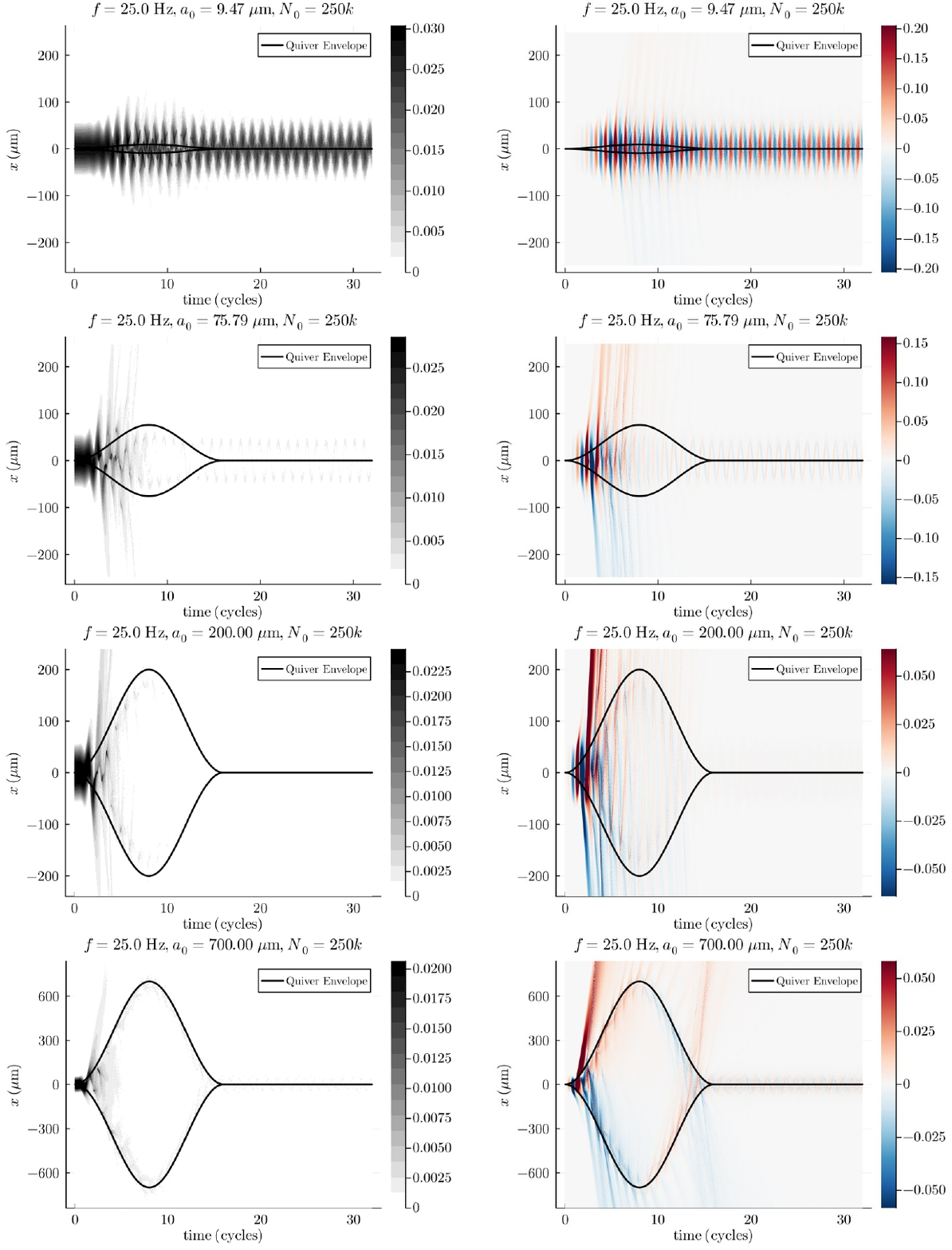}
    \caption{
    Density evolution for $N_0 = 250{,}000$ atoms shaken at $25\,\mathrm{Hz}$ for four
    amplitudes: $a_0 = 9.47\,\mu$m (top), $75.79\,\mu$m, $200\,\mu$m, and $700\,\mu$m
    (bottom). 
    }
    \label{fig:25Hz_densities}
\end{figure}

\begin{figure}[H]
    \centering
    \includegraphics[width=\linewidth]{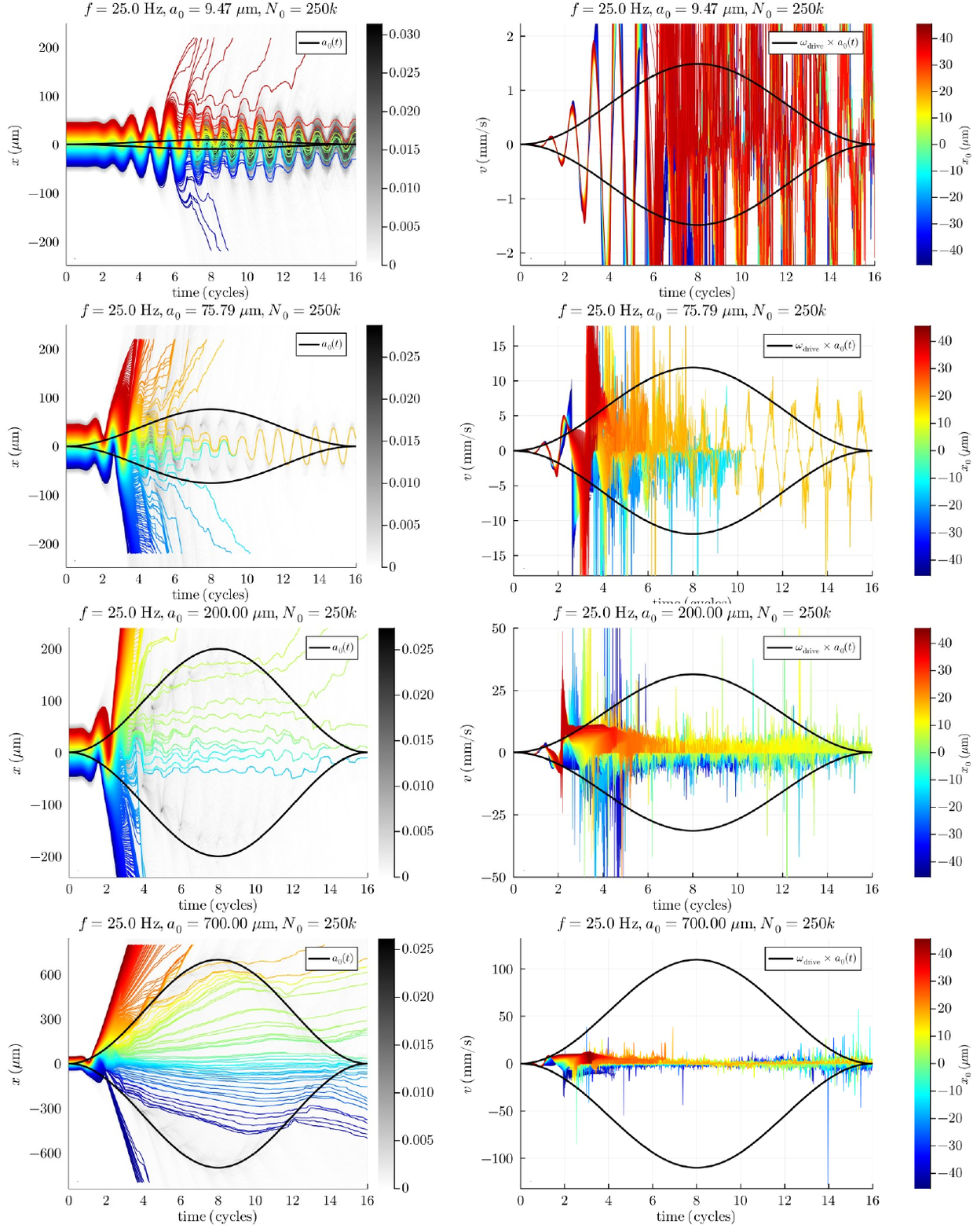}
    \caption{
    Bohmian trajectories (left column) and corresponding velocities (right column) for
    $N_0 = 250{,}000$ atoms shaken at $25\,\mathrm{Hz}$. 
    }
    \label{fig:25Hz_Bohm}
\end{figure}

At the smallest amplitude considered, $a_0 = 9.47\,\mu$m (top rows of
Figs.~\ref{fig:25Hz_densities} and \ref{fig:25Hz_Bohm}), the condensate follows the trap
motion closely and undergoes large‑amplitude sloshing. The density evolution shows noticeable
shape deformation at each half‑cycle, and Bohmian trajectories reveal repeated surface‑layer
ejection whenever the trap reverses direction. This behavior contrasts with the 40\,Hz case,
where the same amplitude produced only weak excitation and negligible loss.

At $a_0 = 75.79\,\mu$m (second rows), the trap motion becomes weakly supersonic, and the
condensate no longer follows the trap as smoothly. The density develops steeper gradients and
more pronounced deformation, and the leading ramp‑up of the envelope produces a strong burst
of loss. During the back half of the pulse, the Bohmian trajectories begin to fall slightly
short of the instantaneous velocity envelope: the atoms attempt to follow the trap but
consistently lag behind it, reflecting the onset of non‑adiabatic, weakly supersonic forcing.
Although some trajectories remain bound after the pulse, loss persists throughout the drive,
and the overall loss fraction is significantly larger than in the 40\,Hz case at comparable
amplitudes.

At $a_0 = 200\,\mu$m (third rows), the trap motion is fully supersonic, and the density
evolution is dominated by strong non‑adiabatic forcing. A substantial fraction of the atoms is
ejected on the leading edge of the pulse, but after several cycles the surviving trajectories
begin to settle into a regime where their average velocities remain near zero. This occurs
because the trap moves faster than the atoms can respond, so the trajectories no longer
attempt to follow the instantaneous trap position and instead begin to experience the slowly
varying, time‑averaged potential. This marks the slow onset of stabilization, consistent with
the remaining‑fraction curve (Fig.~\ref{fig:remaining_fraction}). The stabilization here is
qualitatively similar to the early stabilization seen at 40\,Hz, but it develops more slowly
and at larger amplitudes due to the reduced driving frequency.

At the largest amplitude explored, $a_0 = 700\,\mu$m (bottom rows), the trap motion is
strongly supersonic, and stabilization is fully developed. The Bohmian trajectories show only
small deflections from cycle to cycle, and their instantaneous velocities remain far below the
velocity envelope throughout the pulse. The atoms are effectively decoupled from the rapid
trap motion: the trap moves too quickly for the condensate to track it, and the trajectories
remain confined by the time‑averaged potential with drift velocities that oscillate around
zero. This produces the clearest example of stabilization in the slow‑driving regime. The
current‑density plots reveal a notable distinction between the 200 and 700\,µm cases: at
200\,µm, the flow is still strongly influenced by the instantaneous trap position, with
currents propagating along the bifurcated minima but retaining signatures of the trap’s rapid
oscillation; at 700\,µm, the time‑averaged potential dominates, with currents following the
slowly varying envelope and remaining concentrated near the bifurcated minima.

Relative to the fast‑driving regime at 40\,Hz and the same atom number, the slow‑driving
dynamics exhibit several key differences. Loss at small amplitudes is dramatically larger at
25\,Hz because the condensate follows the trap motion more closely, leading to strong
sloshing and repeated surface ejection. The stabilization threshold shifts upward: amplitudes
that produced strong stabilization at 40\,Hz (e.g.\ $a_0 \sim 75\,\mu$m) do not stabilize the
condensate at 25\,Hz. In the slow‑driving case the trap motion does not become supersonic
until much larger amplitudes, so the system must ride the shaking envelope to higher values
before the time‑averaged potential becomes effective. This extended climb up the envelope
creates a longer window for loss compared to higher‑frequency driving, where the supersonic
regime is reached earlier in the pulse. As a result, stabilization at 25\,Hz emerges only
after several cycles and only at large amplitudes ($\gtrsim 200\,\mu$m), with the
time‑averaged potential becoming dominant only at very large amplitudes ($700\,\mu$m),
whereas at 40\,Hz it dominated already at moderate amplitudes. \textcolor{black}{This behavior matches that of the experimental data depicted in Fig. 3b of the main text, where the frequency-dependent stabilization threshold is demonstrated through measurements of surviving fraction.} These differences highlight
the essential role of the driving frequency relative to the trap frequency in determining
whether the condensate experiences excitation, loss, or stabilization.

\section{From hydrodynamic scales to single-particle scales}\label{sec:one_atom}

Having characterized the response of large condensates under strong shaking, we now turn to the opposite limit: a single trapped atom.  
Moving from atom numbers $N_0 = 40{,}000$--$250{,}000$ to $N_0 = 1$ (or equivalently many atoms with $g\to 0$) replaces the collective hydrodynamic description with purely single-particle dynamics.  
In the many-body regime, the condensate behaves as a compressible fluid whose response is governed by the speed of sound $c$ and the Thomas--Fermi radius $a_{\mathrm{TF}}$, which together determine how quickly density can redistribute and how far the cloud extends spatially.

In the single-particle limit, these hydrodynamic quantities no longer apply.  
Instead, the relevant dynamical scales are set by the properties of the lowest bound state of the external potential.  
The natural length scale is the harmonic-oscillator length
\begin{equation}
    a_\text{HO} = \sqrt{\frac{\hbar}{m\omega_0}},
\end{equation}
and the corresponding velocity scale is the characteristic bound-state velocity
\begin{equation}
    v_{\mathrm{bound}} \sim \omega_0 a_\text{HO}.
\end{equation}
These replace the many-body quantities $a_{\mathrm{TF}}$ and $c$, respectively.

Despite the change in microscopic origin, the structure of the stabilization criteria remains closely analogous.  
For large condensates, strong distortion required (i) supersonic trap motion, $v_{\max} \gtrsim c$, and (ii) shaking amplitudes exceeding the Thomas--Fermi radius, $a_{\mathrm{drive}} \gtrsim a_{\mathrm{TF}}$.  
In the single-atom regime, the corresponding conditions become
\begin{equation}
    v_{\max} \gtrsim v_{\mathrm{bound}}, \qquad
    a_{\mathrm{drive}} \gtrsim a_\text{HO}.
\end{equation}
The first condition expresses that the trap moves faster than the atom can dynamically respond, just as supersonic motion outpaced the propagation of density waves in the condensate.  
The second condition states that the trap displacement exceeds the intrinsic spatial extent of the bound state, mirroring the requirement that the shaking amplitude exceed $a_{\mathrm{TF}}$ in the many-body case.

Thus, while the underlying physics shifts from hydrodynamic flow to single-particle kinematics, the qualitative structure of the stabilization mechanism persists.  
In both limits, stabilization emerges when the external drive outruns the system's intrinsic ability to reconfigure—whether that ability is set by the speed of sound in a fluid or by the bound-state velocity of a single atom.

Periodic driving can stabilize a single bound particle once the external forcing exceeds the intrinsic spatial and dynamical scales of the bound state. In our notation, this occurs when the shaking amplitude becomes comparable to or larger than the harmonic‑oscillator length, $a_{\mathrm{drive}} \gtrsim a_\text{HO}$, and when the maximum drive‑induced velocity exceeds the characteristic bound‑state velocity, $v_{\max} \gtrsim v_{\mathrm{bound}}$. Closely related amplitude‑ and velocity‑based stabilization criteria appear in strong‑field atomic physics, where the quiver excursion and quiver velocity must exceed the characteristic orbital radius and Kepler velocity for ionization to be suppressed \cite{delone_atomic_1995,volkova_ionization_2011}. Experimental searches for this regime have also been carried out in superintense laser fields \cite{de_boer_indications_1993,de_boer_adiabatic_1994}. Similar behavior is known in driven Rydberg atoms, where long‑lived, nondispersive Floquet states form when the microwave‑induced quiver excursion matches the classical orbital radius and the drive frequency locks to the Kepler motion. In that setting, the stabilization mechanism is understood as confinement within a nonlinear resonance island of the classical phase space, and the same amplitude‑ and velocity‑based inequalities govern the existence of localized, non‑spreading electronic wave packets \cite{buchleitner1995nondispersive,buchleitner2002non}. 

Figures~\ref{fig:1_atom_density} and \ref{fig:1_atom_bohm} show the current densities and
Bohmian trajectories for a single trapped atom driven at eight shaking amplitudes,
$a_0 = 18.95$, $75.79$, $94.74$, $104.21$, $113.68$, $123.16$, $161.05$, and
$180~\mu\mathrm{m}$. As discussed earlier, the relevant single-particle scales are the
oscillator length $a_\text{HO}$ and the bound-state velocity $v_{\mathrm{bound}}$, while
the drive introduces the shaking velocity $v_{\max} = a_0 \omega_0$. For all
eight amplitudes we have $a_0 \gg a_\text{HO}$ and $v_{\max} \gg v_{\mathrm{bound}}$,
yet at the smallest amplitudes the shaking is still insufficient to bifurcate the
time-averaged potential. The atom therefore remains localized in a single moving minimum,
and localization occurs before bifurcation in the single-particle limit.

At the smallest amplitude, $a_0 = 18.95~\mu\mathrm{m}$, the atom remains tightly confined
throughout the drive. The current densities show a sharply localized packet oscillating
between the instantaneous turning points $\pm a_0(t)$, and the Bohmian trajectories form a
narrow bundle that closely follows the trap motion. The motion is essentially point-like,
with good following of the trap and negligible loss, consistent with
Fig.~\ref{fig:remaining_fraction}.

At $a_0 = 75.79~\mu\mathrm{m}$ the atom is strongly driven but not ejected. The Bohmian
trajectories show that the packet no longer reaches the full excursion $\pm a_0(t)$; instead
the trajectory bundle lags behind the shaking and remains concentrated near $x \approx 0$
with small oscillations. The current density shows clear oscillations after the shaking
envelope has gone to zero, indicating that the atom has absorbed energy and is left in an
excited bound state once the drive is switched off.

A small further increase to $a_0 = 94.74~\mu\mathrm{m}$ is sufficient not only to excite but
also to eject part of the wavepacket. Both the current densities and the Bohmian
trajectories show loss occurring between approximately 12 and 14 cycles, in agreement with
the dip in Fig.~\ref{fig:remaining_fraction}. Interestingly, some trajectories remain bound
afterwards. The Bohmian plot shows that trajectories born on the upper (positive-$x$)
portion of the initial density are likely to remain trapped, while those starting on the
lower (negative-$x$) portion are more likely to be ejected. This is sensible: ejected
trajectories move downward, and Bohmian trajectories cannot cross, so once the lower part of
the time-averaged trap has flattened the downward-moving trajectories escape while the
upward-moving ones remain confined.

At $a_0 = 104.21~\mu\mathrm{m}$ the remaining fraction increases rapidly. This is clearly
visible in both the current densities and the Bohmian trajectories: some loss still occurs,
but less than at $94.74~\mu\mathrm{m}$. The early-time motion is qualitatively similar to
the previous case up to about eight cycles, but small differences in the trajectory
thereafter are enough to reorient the motion and ensure better bound-state capture during
the ramp-down. For $\beta = a_0/w < \beta_c$ the time-averaged potential retains an
attractive region near the trap center that focuses the trajectory bundle and drives similar
outcomes for all trajectories. As the time-averaged potential flattens and a central barrier
forms at larger amplitudes, trajectories near $x \approx 0$ become increasingly defocused.

Increasing the shaking amplitude to $a_0 = 113.68~\mu\mathrm{m}$ orients the trajectory
bundle so that it strikes the center of the time-averaged barrier. The current densities and
Bohmian trajectories show that comparable portions of the wavepacket are deflected toward
the upper and lower bifurcated minima. This scattering process off the central barrier fans
out the trajectories and leads to a local minimum in Fig.~\ref{fig:remaining_fraction}, as
many trajectories spray over the walls of the potential. The behavior is similar to what is
observed in the many-atom condensate, although there the fan-out is enhanced by $s$-wave
interactions.

At $a_0 = 123.16~\mu\mathrm{m}$ the loss is reduced. The trajectories now pass predominantly
through the lower half of the bifurcated potential and are tightly focused and oriented in a
way that favors recapture. The current densities show a more coherent flow pattern, and the
remaining fraction increases accordingly.

Further increases in shaking amplitude to $a_0 = 161.05~\mu\mathrm{m}$ again orient the
trajectories toward the central barrier, producing a near-equal splitting between the two
time-averaged minima and increased loss. At the largest amplitude considered,
$a_0 = 180~\mu\mathrm{m}$, the trajectory bundle is reoriented so that it passes through the
upper minimum, leading to a slight reduction in loss relative to $161.05~\mu\mathrm{m}$.
Across all amplitudes, the single-particle dynamics show strong sensitivity to the shaking
parameters, consistent with the oscillatory structure of Fig.~\ref{fig:remaining_fraction}.

We operate in a frequency range that is fast enough that the trajectories do not acquire
significant net drift momentum on average, yet not so fast that the phase of the driving
field becomes irrelevant. As a result, the behavior near the cycle-averaged barrier remains
highly sensitive to the shaking amplitude. For certain amplitudes, the incoming trajectory
bundle strikes the nearly symmetric time-averaged barrier in a correspondingly symmetric
manner, sending nearly equal fractions of the trajectories toward the upper and lower
minima. For nearby amplitudes, however, small phase-dependent effects persist, and the beam
can strike the barrier with a slight asymmetry that preferentially deflects the trajectories
upward or downward. Although the trajectories share similar qualitative motion before
encountering the central barrier, this regime is dynamically unstable: small perturbations in
the incoming bundle lead to substantial differences in the post-collision deflection. This
sensitivity to initial phase and amplitude produces the oscillatory loss structure observed
in Fig.~\ref{fig:remaining_fraction}. In contrast to the condensate, the single atom
propagates as a tightly collimated beam that lacks bulk pressure; as a result, it is far
more sensitive to these small perturbations in the unstable barrier region and exhibits
sharper, more phase-dependent deflections. 

\begin{figure*}[h!]
    \centering
    \includegraphics[width=\linewidth]{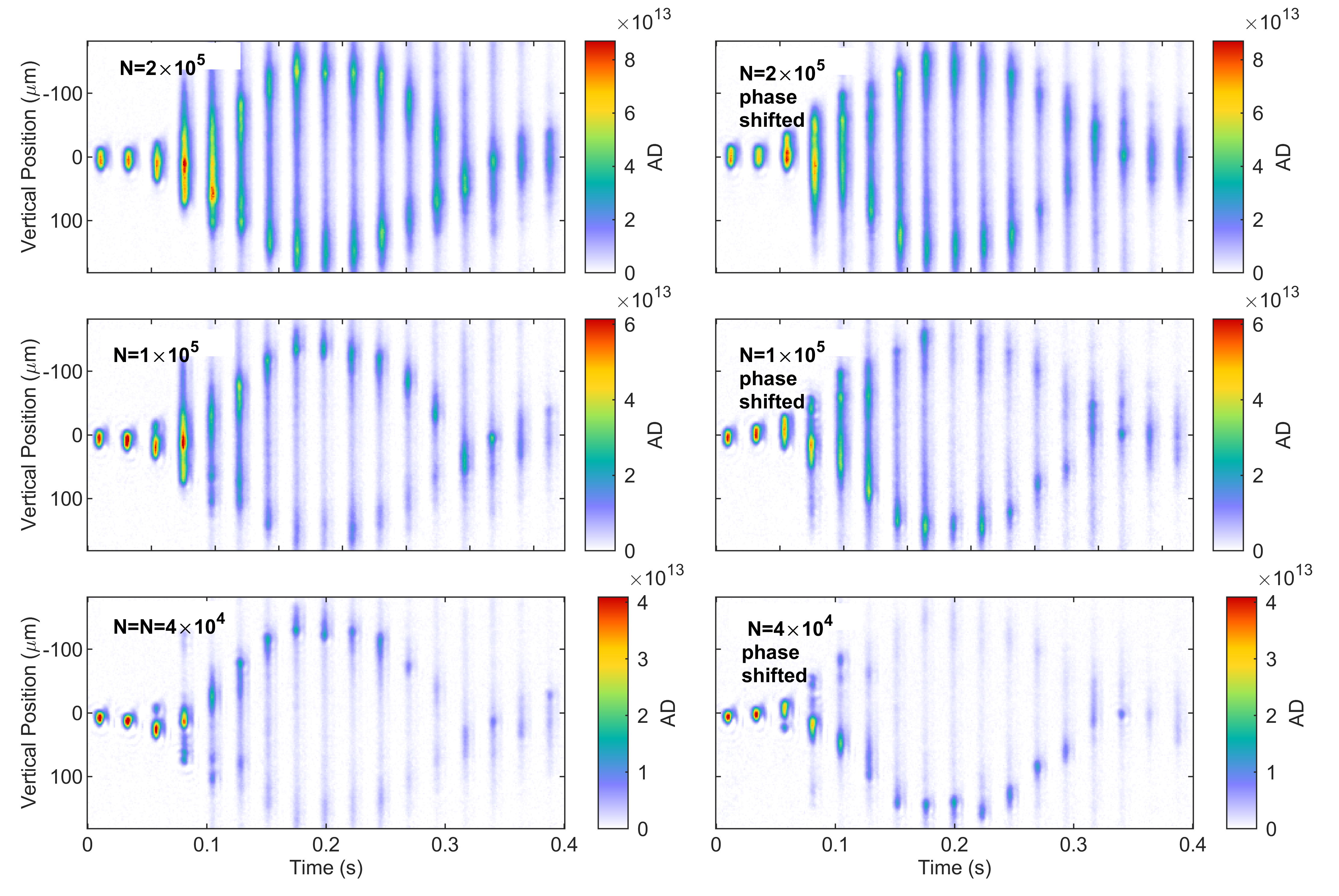}
    \caption{Time-resolved evolution of the density for varied atom numbers. In all cases, the condensate is subjected to a 16-cycle, 40 Hz pulse of amplitude $180\mu\text{m}.$ The condensate is imaged \textit{in situ} after every cycle. Data in the left column follows the same sequence but with a $\pi$-shifted pulse.}\label{fig:VariedAtomNumberODs}
\end{figure*}

\textcolor{black}{Experimental condensate densities in Fig.~\ref{fig:VariedAtomNumberODs} result from 
$N_0 = 4\times 10^4$, $1\times 10^5$, and $2\times 10^5$ shaken atoms. 
\textcolor{black}{The ${}^{84}$Sr-${}^{84}$Sr interaction exhibits a scattering length of $a_s = 123a_\text{B}$, where $a_\text{B}$ is the Bohr radius, giving rise to a mean-field interaction strength $gN_0$, with $g \approx4\pi\hbar^2 a_s/(ma_\text{HO}^2)$ for the trapped condensate. While $a_s$ cannot be tuned for $^{84}\text{Sr}$, the interaction strength can be coarsely varied by varying the total size of the condensate, $N_0$. }
For small interaction strength $gN_0$, the condensate exhibits a clear asymmetry in the bifurcated flow: the two minima of the time‑averaged potential acquire unequal populations, and a $\pi$ phase shift of the drive produces a mirror‑reflected density profile under $x \rightarrow -x$. 
This behavior is most evident for $N_0 = 4\times 10^4$, where the density preferentially occupies a single branch of the bifurcated potential. At larger interaction strength $gN_0$, the flow becomes substantially more balanced. 
For $N_0 = 2\times 10^5$, the densities driven by the electric‑field phase and by a $\pi$‑shifted field are nearly identical, aside from a brief early‑time tendency of the cloud to follow the instantaneous trap motion before stabilization sets in. 
The resulting bifurcated densities are therefore much more symmetric, reflecting the stronger mean‑field repulsion that redistributes population between the two branches. In physical terms, small $gN_0$ corresponds to a tightly focused condensate that can localize in a single minimum, while larger $gN_0$ produces a more diffuse cloud that naturally spreads across both minima in comparable proportion.
}

Taken together, these results illustrate how the same amplitude- and velocity-based
stabilization criteria manifest differently in the two limits. In the condensate, the
conditions $v_{\max} \gtrsim c$ and $a_{\mathrm{drive}} \gtrsim a_{\mathrm{TF}}$ reflect a
collective, pressure-supported response that redistributes density across both minima of the
time-averaged potential. Loss in this regime is observed primarily during the ramp-up
portion of the pulse, and the bulk pressure generated by many neighboring trajectories in
the collisional fluid tends to distribute atoms more evenly than in the non-interacting
case. In the single-particle regime, the corresponding inequalities
$v_{\max} \gtrsim v_{\mathrm{bound}}$ and $a_{\mathrm{drive}} \gtrsim a_\text{HO}$
govern a purely kinematic response: the atom remains localized but becomes increasingly
excited and broadened as the drive strength grows, with loss occurring primarily during the
ramp-down when the trap slows. These length- and velocity-scale criteria describe the
ability of the particle to follow the moving trap and are conceptually distinct from the
geometric bifurcation condition $\beta \gtrsim \beta_{\mathrm{c}}$, which determines when
the time-averaged potential itself splits into two minima. In fact, once bifurcation occurs,
the resulting double-well structure tends to delocalize wavepackets far more strongly than
the time-averaged single-well regime. Thus the qualitative structure of the stabilization
mechanism persists across the two regimes, even though the underlying physics shifts from
hydrodynamic flow to single-particle bound-state dynamics. \textcolor{black}{Additional classical and quantum phase-space analyses of stable single-particle Kramers--Henneberger orbits and their associated scars in the low-frequency limit are presented in Ref.~\cite{floriani2024scars}}.

\begin{figure}[H]
    \centering
    \includegraphics[width=\linewidth,
    trim=0cm 0.5cm 0cm 0cm, clip]{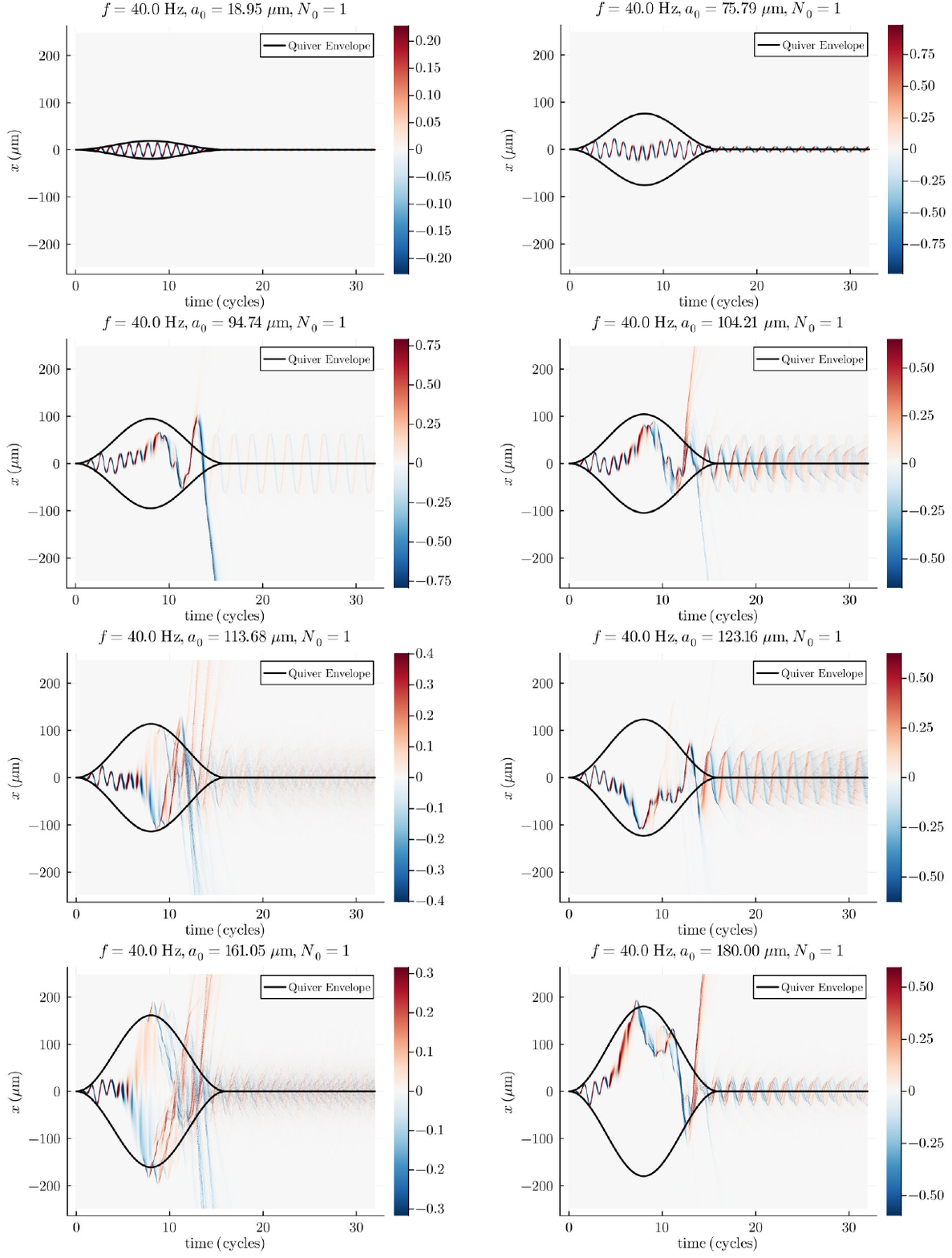}
    \caption{Current densities for a single atom driven at shaking amplitudes
    $a_0 = 18.95$, $75.79$, $94.74$, $104.21$, $113.68$, $123.16$, $161.05$, and
    $180~\mu\mathrm{m}$. The current is plotted in oscillator units, and the positive and
    negative shaking envelopes $\pm a_0(t)$ are overlaid as solid curves.}

    \label{fig:1_atom_density}
\end{figure}

\begin{figure}[H]
    \centering
    \includegraphics[width=\linewidth,
    trim=0cm 2.5cm 1.25cm 0cm, clip]{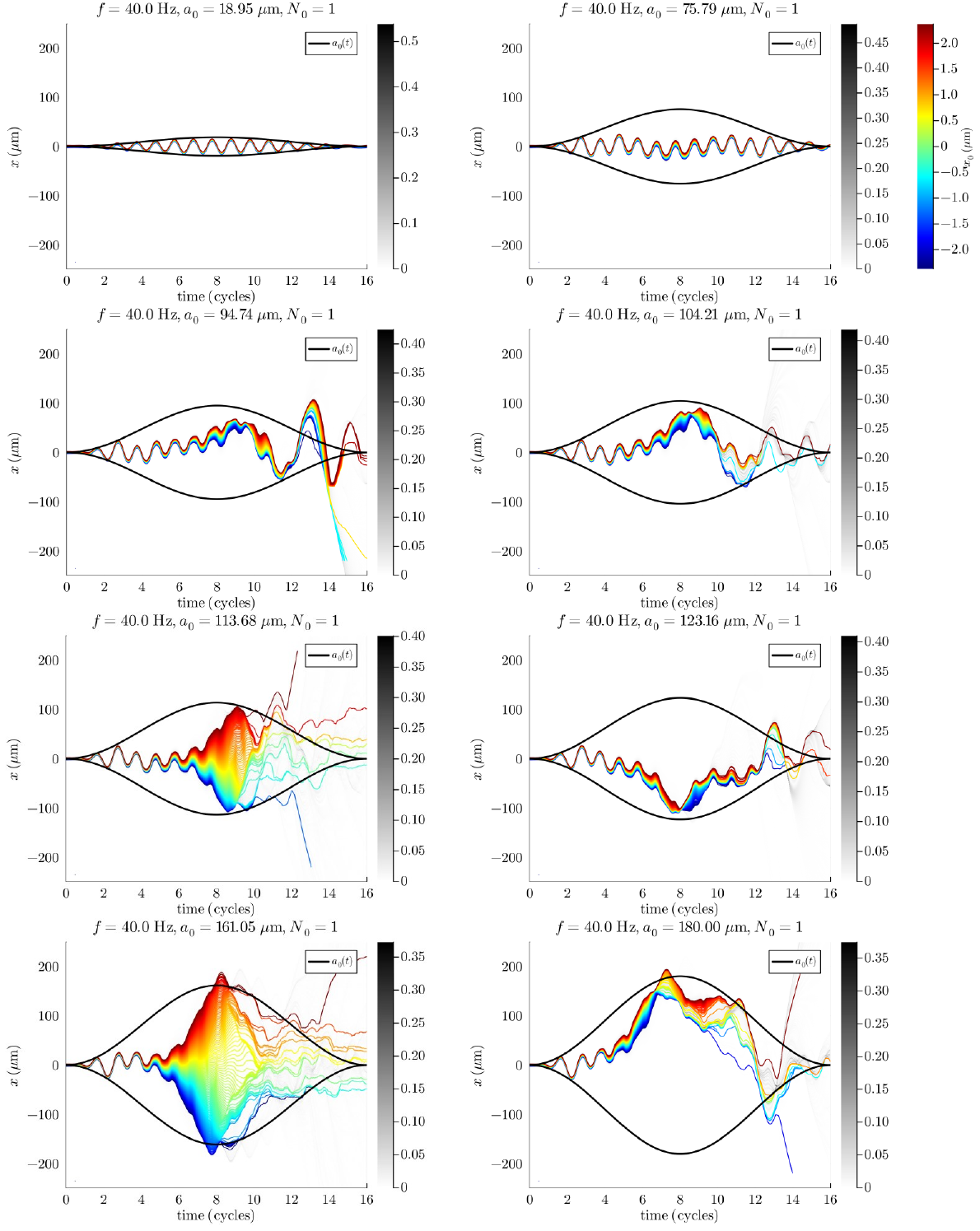}
    \caption{Bohmian trajectories for a single atom driven at shaking amplitudes
    $a_0 = 18.95$, $75.79$, $94.74$, $104.21$, $113.68$, $123.16$, $161.05$, and
    $180~\mu\mathrm{m}$. The trajectories are shown on top of the density profile (in oscillator units), with the positive and
    negative shaking envelopes $\pm a_0(t)$ overlaid as solid curves.}

    \label{fig:1_atom_bohm}
\end{figure}

\section{Broader Context: Emulation of Strong-Field Phenomena}\label{sec:strong-field_emulation}

Our experiment is most naturally viewed as an analog simulator for many-body systems whose
collective dynamics can be captured by a single complex-valued order parameter. In the
electronic context, this corresponds to the many-electron asymptotic limit, where a
Boson-like field $\psi(\mathbf{r},t)$ plays a role analogous to that used in orbital-free
density functional theory and in the statistical theory of atoms
\cite{Kirzhnits_Field_Theoretical_1967,popov1987functional,lundqvist2013theory,mi2023orbital}. Such scalar models reproduce
leading-order properties of heavy electronic systems---including their density profiles
\cite{feynman1949equations} and smoothed, shell-averaged response characteristics, such as
the Thomas--Fermi predictions of Ball, Wheeler, and Fireman for high-frequency
photoabsorption \cite{ball1973photoabsorption} and the collective optical response of metal
clusters \cite{domps1998time}. These approaches inherently smear out densities as they average over fine single-particle
features, whose accurate description requires fermionic, orbital-resolved physics; capturing
such effects would require evolving a full set of $N_0$ spin (or $N_0/2$ spinless) Kohn--Sham
orbitals $\{\psi_j(\mathbf{r},t)\}$ \cite{kohn1965self,zangwill1980density,runge1984density} which is beyond the capabilities of a Bosonic simulator.  

In the Thomas--Fermi description used by Ball, Wheeler, and Fireman, the electrons in a
large-$Z$ atom respond collectively through the TF pressure law rather than through the
orbital motion of individual electrons. Although these authors do not frame their analysis
in terms of a “sound speed,” the TF compressibility naturally defines a collective velocity
scale that governs the high-frequency response of the dense electron gas. In this sense, the
conditions for stabilization in a strongly driven heavy atom may be controlled by this
collective TF response rather than by single-electron Bohr velocities, providing a natural
electronic analogue to the role played by the condensate sound speed in our experiment.
\textcolor{black}{Such intense, high-frequency driving is speculated to be soon accessible at X-ray free-electron laser
facilities, where single-particle stabilization of hydrogen-like atoms in this regime has been explored in theory work~\cite{boitsov2026quasiperiodicdynamicsnondipolexray}.}

As a concrete prototype of the broader class of collective electronic theories we have in mind, we highlight the Thomas--Fermi--Dirac--von Weizsäcker (TFDW) theory of electrons bound to heavy atoms \cite{lundqvist2013theory}. This orbital-free theory describes electronic response using a single collective coordinate $\mathbf{r}$ through the complex-valued order parameter $\psi(\mathbf{r},t)=\sqrt{\rho(\mathbf{r},t)/N_0}e^{(\text{i}/\hbar)S(\mathbf{r},t)}$ \cite{domps1998time,giannoni1976variational}, which describes how $N_0$ electrons evolve in a Bloch-like hydrodynamic manner \cite{Bloch1933} according to the continuity equation
\begin{equation}
    \frac{\partial}{\partial t}\rho +\nabla\cdot \mathbf{j} = 0
\end{equation}
with current density $\mathbf{j}(\mathbf{r},t)\equiv \rho(x,t)\mathbf{v}(x,t)$ and the Hamilton--Jacobi equation
\begin{equation}
    -\frac{\partial S}{\partial t} = \frac{(\nabla S)^2}{2m_\text{e}} + V_\text{tot}
    \quad \text{with} \quad
    V_\text{tot}\equiv V_\text{TF}+\lambda_\text{vW} V_\text{vW}+V_\text{D}+V_\text{ext}+V_\text{H}.
\end{equation}
This describes the conservative flow of an inhomogeneous electron gas bound to parent nuclei, where $\mathbf{v}(\mathbf{r},t)\equiv \frac{1}{m_\text{e}}\nabla S(\mathbf{r},t)$ is a velocity field along which electrons collectively move with mass $m_\text{e}$. Differentiating with respect to $\nabla$ and rearranging yields the Euler equation
\begin{equation}
    \frac{\partial \mathbf{v}}{\partial t} + (\mathbf{v}\cdot \nabla) \mathbf{v} = -\frac{1}{m_\text{e}}\nabla V_\text{tot}.
\end{equation}

The Thomas--Fermi term $V_\text{TF}\equiv \frac{5}{3}c_\text{TF}\rho^{2/3}$ \cite{thomas1927,fermi1927} (with $c_{\mathrm{TF}} = \frac{3}{10}(3\pi^2)^{2/3}\frac{\hbar^2}{m_{\mathrm e}}$) acts as a bulk pressure potential associated with Pauli exclusion in dense Fermionic systems. The von Weizsäcker potential $V_\text{vW}\equiv -\frac{\hbar^2}{2m_\text{e}}\frac{\nabla^2 \sqrt{\rho}}{\sqrt{\rho}}$ \cite{vonweizsaecker1935} gives the exact single-electron kinetic energy and serves as a gradient correction (surface tension) that smooths the sharp edges of the Thomas--Fermi density; its form is identical to the Bosonic pressure potential introduced earlier. The parameter $\lambda_\text{vW}$ controls the strength of this "quantum pressure" contribution, and the commonly used value $\lambda_\text{vW}=1/9$ originates from Kirzhnits' ($N_0\gg 1$) semiclassical expansion of one-particle Green's functions \cite{kirzhnits1957quantum}.  The external potential $V_\text{ext}$ contains the classical electron--nuclear interaction together with any applied driving field (shaking of our trap). 

The remaining contributions, $V_\text{D}$ and $V_\text{H}$, encode electron--electron interactions. The Dirac potential $V_\text{D}$ represents the local exchange energy arising from anti symmetry of the many-electron Fermionic wavefunction and provides an attractive bulk correction that scales as $\rho^{1/3}$ \cite{dirac1930}, partially offsetting the Thomas--Fermi pressure. The Hartree potential $V_\text{H}$ is the classical electron--electron interaction energy obtained self-consistently from Poisson's equation \cite{lundqvist2013theory}.

When the von~Weizsäcker term is neglected (Thomas-Fermi-Dirac approximation), the TFDW potential can be linearized by writing
\begin{equation}
    \rho(\mathbf r,t)=\rho_0(\mathbf r)+\delta\rho(\mathbf r,t),
\end{equation}
where $\rho_0$ is the equilibrium density and $\delta\rho$ is a small fluctuation.
The local part of the effective potential,
\begin{equation}
    V_{\mathrm{loc}}(\rho)\equiv
    V_{\mathrm{TF}}(\rho)+V_{\mathrm{D}}(\rho)+V_{\mathrm{ext}}(\rho),
\end{equation}
admits the expansion (cf.~Ref.~\cite{zangwill1980density})
\begin{equation}
    V_{\mathrm{loc}}(\rho)\approx
    V_{\mathrm{loc}}(\rho_0)
    +g\delta\rho
    =\Big(
        V_{\mathrm{loc}}(\rho_0)
        -g
        \rho_0
     \Big)
     + g
      \rho.
\end{equation}
In direct analogy with a Bose--Einstein condensate in the Thomas--Fermi limit,
the first term in parenthesis acts as an effective trap, while the term proportional
to $\rho(\mathbf r,t)$ plays the role of a local mean-field interaction with
effective coupling (local compressibility)
\begin{equation}
    g(\mathbf r)\equiv
    \left.\frac{\partial V_{\mathrm{loc}}}{\partial\rho}\right|_{\rho_0},
\end{equation}
which is generally position dependent. The Hartree contribution does not permit such a local expansion, as it is
intrinsically nonlocal and must instead be treated through Poisson's equation.

Motivated by the TFDW example above and its potential simulation with BECs, the same
single--order--parameter framework naturally extends to other many--body systems whose
collective dynamics admit hydrodynamic or mean--field descriptions. Electronic motion occurs
on attosecond scales ($\tau_{\mathrm{el}}\!\sim\!10^{-18}\,\mathrm{s}$) \cite{krausz2009attosecond},
whereas nuclear collective dynamics span zeptosecond--attosecond timescales, from giant
resonances ($\tau_{\mathrm{nucl}}\!\sim\!10^{-22}$--$10^{-21}\,\mathrm{s}$) to large--amplitude
shape evolution ($\tau_{\mathrm{nucl}}\!\sim\!10^{-19}$--$10^{-17}\,\mathrm{s}$)
\cite{bohr1969nuclear,bohr1998nuclear}. While BECs cannot reproduce nuclear forces, aspects
of nuclear collective behavior can be \emph{qualitatively} emulated by shaping the trapping
potential to mimic droplet structure \cite{bethe1968thomas} and by using dynamical shaking
to excite collective modes. In this spirit, the analogy between strong--field tunneling,
excitation, and barrier dynamics in atomic physics and the nuclear processes discussed in
Ref.~\cite{zheltikov2025keldysh} offers a useful conceptual bridge. A closely related
two--body analogue of strong--field stabilization appears in Ref.~\cite{smirnova_molecule_2003},
where rapidly oscillating laser fields reshape the effective Coulomb interaction between two
bare nuclei in a manner directly reminiscent of KH--type stabilization in atomic physics.

Related opportunities arise in paraxial photonic fluids, where optical evolution occurs on
picosecond--nanosecond scales ($\tau_{\mathrm{ph}}\!\sim\!10^{-12}$--$10^{-9}\,\mathrm{s}$) and
where the GPE trap potential maps to a spatially varying refractive-index profile
\cite{carusotto2013quantum,larre2015propagation}. This analogy is particularly relevant to
laser filamentation, in which $g<0$ ($g>0$) captures Kerr self-focusing (plasma defocusing),
producing narrow filaments \cite{couairon2007femtosecond} reminiscent of BEC density
profiles, even though the underlying BEC dynamics evolve on much slower millisecond
timescales ($\tau_{\mathrm{BEC}}\!\sim\!10^{-3}\,\mathrm{s}$). In the photonic-fluid analogy,
strong-field stabilization corresponds to reduced beam spreading when oscillations of the
refractive index along the propagation direction generate an effective guiding potential, as
demonstrated in longitudinally modulated waveguides \cite{szameit2010observation}.

\bibliography{Bib1}
\bibliographystyle{unsrt}